# Dynamics of Saturn's Polar Regions

A. Antuñano[1], T. del Río-Gaztelurrutia[1-2], A. Sánchez-Lavega[1-2], R. Hueso[1-2]

[1]Departamento de Física Aplicada I, E.T.S. Ingeniería, Universidad del País Vasco, Alameda Urquijo s/n, 48013 Bilbao. Spain.

[2]Unidad Asociada Grupo Ciencias Planetarias UPV/EHU- IAA (CSIC). Spain

**Index Terms and Keywords:**

```
5704 Saturn Atmosphere

5754 Saturn Polar regions
```

**Key Points:**

- Velocity and vorticity fields of Saturn's Polar Regions are presented.
- Cloud morphology at North and South poles is studied and compared.
- Zonal wind profiles for the Polar Vortices are retrieved and analyzed.

**Abstract**:

We analyze data retrieved by the Imaging Science System onboard the Cassini spacecraft to study the horizontal velocity and vorticity fields of Saturn's Polar Regions (latitudes 60-90°N in June-December 2013 and 60-90°S in October 2006 and July-December 2008), including the Northern region where the hexagonal wave is prominent. With the aid of an automated two dimensional correlation algorithm we determine two-dimensional maps of zonal and meridional winds, and deduce vorticity maps. We extract zonal averages of zonal winds, providing wind profiles that reach latitudes as high 89.5º in the south and 89.9º in the north. Wind measurements cover the intense polar cyclonic vortices that reach similar peak velocities of 150 ms$^{-1}$ at $\pm$ 88.5º. The hexagonal wave lies in the core of an intense eastward jet at planetocentric latitude 75.8°N with motions that become non-zonal at the hexagonal feature. In the south hemisphere the peak of the eastward jet is located at planetocentric latitude 70.4°S. A large anticyclone (the South Polar Spot, SPS), similar to the North Polar Spot (NPS) observed at the Voyager times (1980-81), has been observed in images from April 2008 to January 2009 in the South Polar Region at latitude -66.1º close to the eastward jet. The SPS does not apparently excite a wave on the jet. We analyze the stability of the zonal jets, finding potential instabilities at the flanks of the eastward jets around 70º and we measure the eddy wind components, suggesting momentum transfer from eddy motion to the westward jets closer to the poles.

# 1. Introduction

Since the arrival of the Cassini spacecraft to Saturn, the Imaging Science System (ISS, [*Porco et al.,* 2004]) has provided a vast number of images that picture Saturn's atmosphere with unprecedented detail [*Del Genio et al.,* 2009]. These images have delivered new insight on the dynamics of Saturn's atmosphere, allowing an improved description of both permanent and transient features. The Cassini ISS images have led to a more detailed knowledge of the zonal wind profile at different levels in the atmosphere, together with its variations in time [*Porco et al.*, 2005; *Vasavada et al.*, 2006; *García-Melendo et al.*, 2011], specially when comparing with the wind profiles derived from the Voyagers flybys in 1980 and 1981 [*Limaye et al.*, 1982; *Sánchez-Lavega et al.*, 2000] and Hubble Space Telescope images from 1996 to 2002 [*Sánchez-Lavega et al.*, 2003, 2004]. The Cassini ISS images have also allowed a high resolution analysis [*García-Melendo et al.*, 2013; *Sayanagi et al.*, 2013a; *Dyudina et al.,* 2013] of one of the most conspicuous phenomena in Saturn, the Great White Spots, with the eruption of the last event late in 2010 [*Sánchez-Lavega et al.*, 2011a] as well as other dynamical phenomena [*Sayanagi et al.,* 2014].

The polar regions of planetary atmospheres have singular behaviors highly conditioned by the planetary rotation rate. In the case of Saturn, a fast rotating planet, a great amount of attention has been drawn by the hexagonal feature approximately centered at 75º planetocentric latitude in the Northern Hemisphere. This feature, that encloses a fast eastward jet, was first observed by Voyager 1 and 2 flybys in 1980 and 1981 [*Godfrey*, 1988], and later on during 1990-1995 by the Hubble Space Telescope (HST) and with ground based telescopes [*Sánchez-Lavega et al.*, 1993; *Caldwell et al.*, 1993]. A big elliptical anticyclone known as North Polar Spot (NPS), of an east-west size of ~7,000-10,000 km was observed just outside the hexagonal feature at the time of the Voyager images in 1980 and 1981 and was still present in ground-based observations in 1995 [*Sánchez-Lavega et al.*, 1997]. The NPS was reddish in color and bright in methane absorption band images [*Hunt and Moore*, 1982; *Sánchez-Lavega et al.*, 1997]. It was proposed that the hexagonal feature was created by the NPS impinging on an intense eastward jet embedded in the hexagon [*Allison et al.*, 1990; *Sánchez-Lavega et al.*, 1997]. Nevertheless when the hexagon was observed again by Cassini, first during Saturn's night in 2007-2008 by the Composite Infrared Spectrometer (CIRS) [*Fletcher et al.*, 2008] and then by the Visual and Infrared Mapping Spectrometer (VIMS) [*Baines et al.,* 2009] and by the ISS [*Sánchez-Lavega et al.*, 2014] the NPS was no longer present. The NPS is therefore not necessary to the presence of the hexagon, and this leaves open the question of what forces the hexagonal wave. An accurate knowledge of the hexagonal jet structure and the role played by smaller scale eddies inside and around it is needed to explore formation scenarios linked to atmospheric instabilities or wave development. The persistence of the hexagon over this period and the remarkable stability of its rotation rate through the polar seasons have led to the suggestion that it is a deeply rooted atmospheric structure whose rotation rate may reveal the internal planetary rotation rate [*Sánchez-Lavega et al.*, 2014] which is currently constrained by measurements of magnetic field [*Sánchez-Lavega,* 2005; *Gurnett et al.,* 2010].

HST images captured between 1997 and 2002, and ISS images after 2004, showed that a fast eastward jet was present at similar latitudes in the southern hemisphere [*Sánchez-Lavega et al.*, 2002], but despite the apparent similarity between the north and south jets, there was neither a hexagonal feature nor a big anticyclone in the southern hemisphere. On the other hand, on its approach to Saturn, Cassini confirmed the presence of a circular feature at the planet's South pole that had been already revealed in images in the thermal infrared captured form Earth [*Orton and Yanamandra-Fisher*,

2005]. High resolution images showed a polar vortex [*Sánchez-Lavega et al.,* 2006] with a dark eye surrounded by fast moving white clouds at -88º planetocentric [*Dyudina et al.*, 2008]. The existence of a North Polar vortex similar to that in the South was first suggested by *Fletcher et al.* [2008] from analysis of thermal data from the night-side North Pole during its winter. Later on in the mission, when Saturn's North pole began to be illuminated by sunlight after the spring equinox, a mirror cyclone to the South Polar vortex was indeed observed in the cloud fields of the northern pole. These two intense polar vortices are circular stable features very different to those found in the atmospheres of Earth, Mars [*Mitchell et al.* 2014], Venus [*Piccioni et al.* 2007*; Luz et al.* 2011*; Garate-Lopez et al.* 2013] and Titan [*Teanby et al.,* 2010], linked to the surface (Earth and Mars at least) or forming under slow planetary rotation (Venus and Titan). Thus, polar vortices on the giant planets (rapidly rotating fluid bodies with deep atmospheres) are of great interest for comparative planetology in view of their formation under very different dynamical and thermo-chemical conditions to those on terrestrial bodies. In addition, Saturn suffers a strong seasonal insolation cycle at the poles [*Pérez-Hoyos and Sánchez-Lavega*, 2006], so the knowledge of the possible influence of insolation on the polar dynamical features becomes relevant to understand their nature.

In this paper, we use high-resolution Cassini ISS images to study the atmospheric dynamics at cloud level of the North and South Polar Regions with planetocentric latitudes between 60º and 90º. We obtain precise wind measurements using a supervised cloud correlation algorithm [*Hueso et al.*, 2009] that allows retrieving vorticity maps of both regions, and we make a comparative analysis of the morphology and velocities of both regions. Although results of the zonal winds of the South Polar Region are already available [*Sánchez-Lavega et al.* 2006*; Dyudina et al.* 2008] the current work improves over those measurements describing accurate zonal and meridional motions, and provides a first detailed analysis of the motions in the North Polar Region, which allows a systematic comparison of the morphologies and dynamics of both polar regions.

## 2. Database and measurement methods

### 2.1. Image selection

In this study we have used several sets of high resolution images taken with the Imaging Science System (ISS) onboard the Cassini spacecraft. Information about the instrument, its wide angle camera (WAC), narrow angle camera (NAC), and set of available filters can be found in *Porco et al.* [2004]. We have used images captured in four different periods: a) A set of WAC and NAC images taken in June 2013, using CB2 and CB3 filters for the study of the North Polar Region, b) Two sets of images taken with the WAC and CB2, CB3 filters in October 2006 and December 2008 for the study of the South Polar Region and South Polar Vortex and c) a set of NAC and CB2 and red filters in July 2008 for the study of the south polar vortex. The choice of CB2 and CB3 narrow filters (centered at 752 nm and 939 nm respectively) is due to the fact that they reveal the features of both Polar Regions with the highest contrast, allowing easier feature identification. Both filters show the same structures with similar contrast and are assumed to sense cloud features at the top of the ammonia cloud. The sensed level ranges between ~350 – 750 mbar from detailed radiative transfer analysis [*Pérez-Hoyos et al.,* 2005; *Sánchez-Lavega et al.* 2006; *García-Melendo et al.*, 2009]. While other image filters were available at the time of the observations, they either represent the upper hazes without contrasted features, or the same cloud level as observed in CB2 and CB3 filters but with reduced contrast [*Dyudina et al.* 2008; *Del Genio et al.* 2009]. Our use of NAC images limits to the study of north and south polar vortices, while all other regions, down to ±60º planetocentric latitude, have been studied with WAC images. We have searched for pairs of images that depict the same region of the planet in two different times, with time intervals varying from 60 minutes to approximately 10 hours (one planetary rotation). In Table 1 we summarize the images that were used, together with some of their more relevant characteristics. On some instances a given image has been used in more than one image pair.

[Table 1]

**2.2. Measurement method**

Selected images are navigated using the software PLIA [*Hueso et al.*, 2009]. The software uses information extracted from SPICE C-kernels [*Acton*, 1995] to associate coordinates to each pixel of the image, allowing a manual correction of the navigation via limb-fitting when the limb is visible in the image, and/or adjusting the position of the pole. Once the navigation is corrected, the images are polar projected using an azimuthal equidistant projection [*Snyder,* 1987]. Although this kind of projection does not preserve the area, the maximum area distortion, at ±60º, is ~ 5% and reduces quickly as we approach the pole (at the mean latitude of the hexagon the distortion is only 1%). We use maps including latitudes from 60º to 90º in the North Polar Region and from -60º to -90º in the South Polar Region with a meridional resolution of 0.05º/pixel, again using the software PLIA, while high resolution images of the polar region are projected in maps including latitudes from 85º to 90º in the North, with a meridional resolution of 0.01º/pixel, and from -85º to -90º in the south, with a resolution of 0.02º/pixel. All throughout this work, we use planetocentric latitudes and longitudes in Saturn System III reference frame with angular velocity $\Omega_{III}$ = 810.7939024 º/day [*Desch and Kaiser*, 1981; *Seidelmann et al.,* 2007].

In order to measure wind velocities we assume that the small-scale cloud features resolved in the images follow the wind flow and we track their motions in images pairs separated by a given time interval [*Sánchez-Lavega et al. 2000*]. We use two alternative methods to measure the displacements of the clouds depending on the time interval between images:

a) An automatic two-dimensional brightness correlation algorithm, that allows the experimenter to validate, correct or ignore automatic identifications [*Hueso et al., 2009*]. Due to the fast evolution of cloud features, this algorithm is adequate for images separated by short time intervals up to approximately 120 minutes but it can also be used safely in images separated by a 10 hr interval (close to a planet rotation). This method has been used to examine all image pairs; however, rotation of the cloud features, significant in the inner polar vortices even in images separated by short time-intervals due to their fast rotation, decreases the efficiency of the algorithm. The size of the correlation boxes is chosen to be large enough to allow the experimenter to validate the correlation (boxes too small make visualization of the features difficult and introduces noise in the measurement, while on the other hand larger boxes lead to fewer wind vectors and smoother results, with the corresponding loss of information of small perturbations that may be present in the wind profile). We have used boxes of 23 x 23 pixels for the northern hemisphere, and 25 x 25 pixels for the south hemisphere, with a pixel corresponding to ~50 km in projections from 60º to 90º in both hemispheres and to 10 km and 20 km in high resolution maps of the north and south poles respectively. The validation step resulted in 5% of the candidate wind vectors presented by the software rejected, and 5% of the original vectors corrected. The overall result of the validation step results in coherent maps of wind vectors without obvious outliers that need to be removed afterwards [*Del Genio et al.* 2007].

b) Visual cloud tracking in high-resolution NAC image pairs separated by times of 40-80 minutes where the previous method offered only a few wind vectors for each image pair. This method produced a minor fraction, ~1%, of the total wind measurements, but these measurements covered some of the most difficult regions, such as the periphery of large vortices in the polar regions and the innermost part of the polar vortices.

In the case of the South Polar Region, the original images were bland due to the viewing geometry (the sub-spacecraft point was on average -40ºS) and low illumination of these latitudes (subsolar point ~-8ºS), and they had to be processed to enhance the details. We used a combination of unsharp-mask and high-pass filters to enhance the contrast of the different features in order to measure the highest number of features in all the regions of our study [*Hueso et al.*, 2010]. This was not necessary in the North Polar Region because the original images had very good contrast resulting from the better illumination (subsolar point at 19ºN) and better observing geometry (sub-spacecraft latitude of ~52ºN), without the need for image processing.

There are several sources of error in the determination of the velocities of features. Extensive experiments with the parameters defining the correlation algorithm, i.e. changing the correlation box size and minimum correlation factors, give a typical variation of wind speeds of 3 ms$^{-1}$ in the worst cases and the correlation algorithm is assumed to produce results at least valid to the pixel resolution of original images. The resolution of the images imposes a lower limit to the error of an individual measurement, which also depends on the time interval between images:

$$\sigma(u) = \frac{\text{Image resolution}}{\Delta t} \qquad (1)$$

Another important source of error is navigation error. In the case of images obtained by the WAC camera, uncertainty in navigation is better than half a pixel from fitting an ellipse to the planetary limb. However in very high resolution NAC images where the limb is not visible the navigation uncertainty can be as large as ten pixels, leading to an error in the determination of velocity similar to that of the WAC images when using the same time interval between images. The combination of these errors is different for each particular date and it is presented on Table 2. It can be summarized saying that the mean cloud tracking uncertainty for individual measurements varies on the order of 5-10 ms$^{-1}$. The changing illumination conditions on different dates have an impact in the effective area of the polar regions that can be examined to extract wind measurements, and a lower number of measurements were obtained for the regions poorly illuminated.

[Table 2]

## 3. Results: Two-dimensional structure of wind fields

### 3.1 Morphology of the cloud field

In Figure 1 we present two mosaics showing polar projections of the North and South Polar Regions from 60º to 90º and from -60º to -90º. The north polar projection is built using one image on 14 June 2013 during spring in the northern hemisphere and the south polar projection is built using four images on 3 December 2008, late summer in southern hemisphere. Limb-darkening from individual images have been partially compensated by the use of a Lambert correction and knowledge of the illumination geometry. In fact, the illumination under which the original images were taken was quite different, and resulted in poorer contrast in the images of the southern pole when compared with those of the north. Nevertheless after some contrast enhancement, many similar features between the two hemispheres are apparent. In the mosaic shown in Figure 1, the polar vortex at the South Pole appears brighter, but this is an artifact of the mosaic composition and the use of Lambert correction, which over corrects the latitudes where the illumination is almost tangent, rendering them brighter than they are. When the individual images are examined without limb-darkening correction, the South Polar vortex has a similar aspect to the North vortex, as can be seen in Figure 2.

Both polar vortices extend in latitudes from 88.5º to 90º (north or south) [*Sánchez-Lavega et al.*, 2006; *Fletcher et al.*, 2008; *Dyudina et al.*, 2009; *Sayanagi et al.*, 2013b], and are enclosed by a very fast eastward jet. Figure 2 shows a high resolution polar projection of both poles of Saturn at the same scale to facilitate comparison. The cloud morphology of the images reveals by itself the vortex nature of the polar dynamics (see below). The cloud texture seems richer and more complex on the North, which is more populated with clouds, but this could be an effect due to the different viewing angle conditions combined to the wavelength of the filters.

**[Figure 1]**

**[Figure 2]**

At lower latitudes, both hemispheres display another fast eastward jet, but this time the latitude of the jets in the North and South hemispheres are quite different. These jets, centered at 75.8±0.1ºN in the North and -70.4±0.1ºS in the South, reveal themselves in elongated clouds probably torn apart by the sheared flow. The most striking difference between the northern and southern polar regions is the presence of the well-defined permanent hexagonal feature in the northern hemisphere [*Godfrey,* 1988; *Caldwell et al.*, 1993; *Sánchez-Lavega et al.*, 1993] that shapes the jet at 75.8ºN latitude. This feature is not observed at the South Polar Region, where the symmetric eastward jet to the hexagon follows the latitude circle instead. While there are a few regions of the jet slightly distorted into evanescent quasi-linear features, they do not match an entire or fractional wave number. These quasi-linear features appear in different dates with different characteristics and were noted before by *Vasavada et al.* [2006]. These authors also noted the presence of a low amplitude polygonal distortion of the 60.5ºS jet with an ephemeral 12 wave number structure that was present only on a few dates.

In both hemispheres "puffy clouds" are present at latitudes above and below the eastward jet. These features vary in size, typically up to ~ 300km in the north hemisphere and up to ~ 400km in the south hemisphere. The number of puffy clouds seems to be higher in the North Polar Region than in the South Polar Region. The pattern they form is reminiscent of the cumulus cloud fields that forms on Earth under moist convective instability [*Houze*, 1993].

Finally, singular vortices of different sizes also appear at polar latitudes in both hemispheres. *Vasavada et al.* [2006] present a global census of vortices in Saturn in the South hemisphere from data obtained in 2004 and *Trammel et al.* [2014] present a global census of vortices in both hemispheres from low to high-latitudes without covering the subpolar latitudes. In the polar maps shown in Figure 1 there is a very noticeable anticyclonic vortex at 80.5ºN with a zonal dimension of 18° ± 1° (2,850 ± 160 km) and a meridional dimension of 2.5° ± 0.5° (2,380 ± 480 km). This vortex was first observed in October 2012 and is still visible in Cassini ISS images. Note that the North Polar Spot (NPS) that was discovered at the time of the Voyagers flybys in 1980-1981 and was visible on some ground-based observations up to 1995 [*Sánchez-Lavega et al.*, 1993] has vanished during the North polar winter and is absent from all observations with Cassini. In the South Polar Region, there are two notable singular vortices: 1) a circular anticyclone at -73.9º ± 0.1º with a zonal size of 9°± 1º (2,390 ± 260 km) observed at least from July 2005 to January 2009 and 2) an anticyclone at – 66.1º ± 0.1º with a zonal dimension of ~20º ± 1º (7,800 ± 400 km) and meridional dimension of 5º ± 0.5º (4,800 ± 400 km) observed from April 2008 to January 2009, which we will denote South Polar Spot (SPS) due to its similarity with the North Polar Spot (NPS). A similar feature to this vortex is seen in the South polar map presented by *Vasavada et al*. [2006] corresponding to data from 2004 but its smaller size and different drift do not allow us to conclude that it is the same vortex as the SPS. Other smaller-scale vortices (like the dark vortex at the core of the jet at -70.4ºS) might survive from 2004 to 2008-2009, but there are not enough data to prove that they are the same structures. We cannot thus exclude the possibility that all these vortices are transient features that appear and disappear or evolve in the polar atmospheres in time-scales not covered by the data here presented.

**3.2 Two-dimensional wind fields**

As we described in section 2.3, two different methods have been used to measure wind velocities, two-dimensional brightness correlation [*Hueso et al* 2009], more adequate for image pairs close in time (up to 120 min), and visual cloud tracking, when image pairs separated by several hours are available, since in that case the correlator is not able to identify features due to changes in their morphology. All pairs of WAC images have been analyzed with the brightness correlator, while in the case of images captured by the Narrow Angle Camera we have used both methods with the same pair of images to verify the results, as some of the clouds at the polar vortices move very rapidly and their displacement is not captured by the automatic correlator. For the measurement of the wind vectors of the North Polar region we use five pairs of WAC images and two pairs of NAC images from June 2013, while for the South Polar Region we use eight pairs of WAC images (four from December 2008 and four from October 2006) and one pair of NAC images from July 2008.

Figure 3 shows the density of wind-vector measurements in both hemispheres combining the data for all dates, again using polar azimuthal equidistant projections. The density of measurements is examined in square bins of 6 degrees in latitude which roughly corresponds to physical sizes of 5,690 x 5,690 km. The North Polar Region has a larger number of measurements, with a number of wind vectors per box varying from more than 600 wind vectors in the best cases to just a few wind vectors in the worst cases. In the south, the best covered regions have about 300 measurements, and most relevant regions have 200 vectors per bin or less. This is related to the better viewing conditions of the North polar atmosphere. Additionally, a sector of longitudes in the South polar atmosphere has no measurements due to the absence of a suitable image pair covering that region. The total number of wind measurements is 14,910 wind vectors for the North Polar Region and 7,925 wind vectors for the South Polar Region, both confined between 60º and 90º.

**[Figure 3]**

Figure 4 shows the wind field from the ensemble of measurements and dates. Note that the absence of large-scale features in the South hemisphere and the small time separation of the data used to study the north hemisphere result in homogenous global results. We interpolate the winds into a regular cartesian grid with a step of 285 km (which corresponds roughly to 0.3º in latitude). Since the original data is obtained in a set of points that are not regularly separated the interpolation is done using the

Delaunay triangulation method and bilinear interpolation for the evaluation of the wind data in each new grid point. Finally the data is smoothed with a boxcar average to a spatial scale of 3,800 km (4º in latitude). Panels A and C show the two-dimensional velocity field for the northern and southern hemispheres respectively. One of every twenty five interpolated wind vectors is represented here. Panels B and D zoom into the polar vortices.

[Figure 4]

Figures 5 and 6 show colored contour maps of the mean zonal and meridional velocities. In order to build these figures we created a polar centered x-y grid of 285 km (which corresponds to 0.3º/pixel in latitude), interpolated our measured velocity vectors onto the grid using the previous method, and smoothed the resulting values with a boxcar average to a spatial resolution of ~1º in latitude (1,140 km).

[Figure 5]

[Figure 6]

Figures 4 and 5 show the strong eastward jets at 75.8ºN and 70.4ºS, with the hexagonal feature in the northern hemisphere and its meandering jet, while in the south the jet does not depart significantly from a pure zonal motion, with wind vectors essentially tangent to the latitude circles. The polar jets (or outer rim of the polar vortices) at 88.5ºN and 88.5ºS can also be seen, especially in Figure 5, with velocities reaching ~160-170 ms$^{-1}$ in both hemispheres. It is also apparent that the southern polar jet is wider than its northern counterpart. The eastward jets at lower latitudes, 75.8ºN in the north and 70.4ºS in the south, are slower, with maximum interpolated velocities of the order of 100-110 ms$^{-1}$. We note however that some of our measurements by cloud tracking show features moving with zonal velocities as fast as 120 ms$^{-1}$ at those latitudes, implying a significant dispersion of velocities at the core of these highly sheared jets.

On the other hand, the interpolated meridional velocities in Figure 6 show the hexagonal feature as a prominent set of meridional perturbations ~ ±28 ms$^{-1}$ (Figure 6A) that represent the meandering motion of the hexagon jet, while in the south (Figure 6B) there are no important departures from a purely zonal motion. The North vortex at 80.5ºN and the South Polar Spot (SPS) at 66.1ºS are also prominent in the meridional velocity maps with meridional velocities of 15 ms$^{-1}$ in the north vortex and 35 ms$^{-1}$ in the SPS. This is due to the fact that although figures 4-6 present the ensemble of our data, at the particular location of the vortices we have filtered out of these maps data from dates other than 14 June 2013 in the North and or 3 December 2008 in the South. Small scale inhomogeneities in Figures 5 and 6 and the overall velocity gradient observed in Figure 6B represent velocities of less than 10 ms$^{-1}$, that is, of the order of our measurement error and, therefore, are not significant. Finally, let us point out that although most regions covered by these maps have local dispersions of zonal velocities of less than 10 ms$^{-1}$, the core of the eastward jets and the highly sheared polar vortices have significant dispersions of velocities of the order of 15-20 ms$^{-1}$.

## 4. Results: Zonal wind profiles

Except for the hexagon jet, motions in the polar regions are dominantly zonal (eastward or westward). To determine mean zonal wind profiles for the north and south polar regions, we have averaged individual zonal velocities in latitudinal bins of 0.4º in intervals of 0.2º of latitude in the North Polar Region, and in bins of 0.6º in intervals of 0.3º in the South Polar Region. This way, each measurement lies in two different bins producing a smooth wind profile. The choice of a larger bin in the south is due to the smaller number of wind vectors measured in that region. In both cases, the highest latitudes (from 89º to the pole in both hemispheres) are studied with bins of 0.2º in intervals of 0.1º since the data comes from the highest resolution observations. The result of this averaging process, together with the standard deviation of each latitudinal bin is given in table A1 for the north and table A2 for the south. Figure 7 shows the results.

The standard deviation in each latitudinal bin is of the order of the navigation error for the westward jets at both polar regions and slightly larger, in the range 10-20 ms$^{-1}$ in the eastward jet at 70ºS. However, it is larger at both polar vortex jets, due to the small number of measurements at those latitudes, and at the eastward jet associated with the hexagon in the north, due to the fact that we have ignored that the mean winds are not strictly zonal in that region, since they follow the meandering hexagon.

The most polar eastward jets at the north and south poles reach similar velocities of $u$=140 ± 15 m/s in the north and $u$= 157 ± 12 m/s in the south. Both jets are clearly asymmetric in shape, with their peaks at similar latitudes but different FWHM, 4.5 ± 0.1º in the north and 5.9 ± 0.1 º in the south. They are stronger than the next eastward jets at lower latitudes. The jet containing the hexagon reaches $u$= 104 ± 15 m/s while its counterpart in the south reaches $u$= 87 ± 12 m/s, and is approximately 5º degrees farther from the pole than its north counterpart.

**[Figure 7]**

To analyze the average flow of the hexagon jet taking into account its meandering motion, we have extracted a mean profile by averaging regions not of equal latitude but of equal perpendicular distance to the peak of the jet. In order to do so, we have calculated mean zonal ($\bar{u}$) and meridional ($\bar{v}$) velocity averaged over all observations in bins of 8º longitude in steps of 4º, and 0.5º latitude in steps of 0.25º, covering the region from 60º to 83º. In this way, we obtain 90 different latitudinal profiles, one for each longitudinal bin. We deduced the latitude of the jet maximum for each longitudinal bin, performing for each longitude a sixth grade polynomial fit of the modulus of the velocity versus latitude. Figure 8A shows the values of the planetocentric latitudes of the peak of the hexagonal jet against longitudes in System III. We see that the deviation of the latitude of the maximum from its mean value fits a sinusoidal function with zonal wave-number 6 as

$$\varphi_{max}(\lambda) = \varphi_0 + \Delta\varphi_0 \sin\left(\frac{2\pi}{\Lambda}(\lambda - \lambda_C)\right), \qquad (2)$$

where $\varphi$ and $\lambda$ are the latitude and longitude, and the mean latitude is $\varphi_0 = 75.79^0 \pm 0.01º$, the amplitude of the oscillation is $\Delta\varphi_0 = 0.49^0 \pm 0.01º$, the wavelength $\Lambda = 59.7º \pm 0.2º$ and the phase shift $\lambda_c = -6.0º \pm 0.7º$, related to the orientation of the hexagon. This wave is represented on top of a polar projection of the north region in figure 8B. Thus, we see that the hexagonal feature is a sinusoidal wave that is observed as a hexagon because the oscillation amplitude is just right to form a structure that looks almost linear. A hexagonal wave of larger amplitude would result in a clearly wavy feature without the quasi-linear sides of the hexagon, while a lower amplitude would render the hexagonal perturbation unobservable.

**[Figure 8]**

Once the latitude of the jet's peak is known, eqn. 2 allows us to define a new zonal coordinate that is centered at the peak of the zonal velocity wind, with the x axes parallel to the jet motion and the y axes perpendicular to it. Using this new coordinate the average profile is slightly narrower and 5 ms$^{-1}$ faster, with a smaller standard deviation.

Finally, to further compare the jets at 75.8ºN and 70.4ºS, we have fitted them to Gaussian function of the type $u = u_0 \exp(-bR^2(y-y_0)^2/2u_0)$, where $R$ is the local planetary radius, $u_0$ the peak velocity and $b$ the jet curvature [*Allison et al.*,1990]. These fits show that the hexagon jet is ~25 ms$^{-1}$ faster and ~400 km narrower than its southern counterpart. Details of the jets and their fits are shown in Figure 9.

[Figure 9]

The combination of high velocities and extreme latitudes close to the poles requires the polar vortices to be in gradient wind balance,

$$\frac{u^2 \tan\varphi}{R_p(\varphi)} + fu = -\frac{1}{\rho}\frac{\partial P}{\partial y} \qquad (3)$$

This equation is valid for lower latitudes, where the centrifugal term vanishes, and can be used to integrate the pressure field over the polar latitudes. Choosing an atmospheric density $\rho$ = 0.12 Kg m$^{-3}$ at 0.5 bar altitude level, a Coriolis parameter $f$ = 3.28x10$^{-4}$ s$^{-1}$, a local radius $R$ = 54,365 km, $\varphi$ = 88.7º and $u$ = 150 ms$^{-1}$ we find that the pressure gradient is $\partial P/\partial y = 7.4 \times 10^{-3}$ Kg s$^{-2}$m$^{-2}$ or $7.4 \times 10^{-2}$ mbar/km, which is approximately half the typical value of a hurricane on Earth, 0.17 mbar/km [*Emanuel, 1994*]. In figure 10 we show a detail of the north and south zonal wind profiles in the polar region, together with the pressure difference with the pole. We find a pressure difference between the pole and the peak of the polar jet of ~ -65 and -50 mbar in the north and south polar vortices respectively. Therefore, both polar cyclones are low-pressure cores, and they are warmer than their surroundings, as inferred from CIRS and

VIMS data [*Fletcher et al.,* 2008]. The warmer temperatures are probably related to subsidence motions inside both polar vortices and the thermal wind equation predicts a drop with altitude in the wind intensity of the vortices. *Sánchez-Lavega et al.* [2006] present a dynamical analysis of the South polar vortex with an estimated global subsidence of 1.4 ms$^{-1}$ and decreasing winds with altitude of 10 ms$^{-1}$ per scale height. The same rough numbers can be applied to the north cyclone on the basis of the similar wind field, size and thermal structure. The detailed structure of the vortex in Figure 1 shows a spiraling pattern of clouds towards the cyclone that agree with a global subsidence pattern.

[Figure 10]

## 5. Vorticity and vorticity gradients

### 5.1 Two-dimensional vorticity maps

The vertical component of the relative vorticity of a fluid in spherical coordinates is given by [*Sánchez-Lavega,* 2011b]

$$\xi(\lambda,\varphi) = \frac{1}{R\cos\varphi}\frac{\partial v}{\partial \lambda} - \frac{1}{R}\frac{\partial u}{\partial \varphi} + \frac{u}{R}\tan\varphi, \qquad (4)$$

where *u* and *v* are zonal and meridional velocities, $\varphi$ the planetocentric latitude, $\lambda$ the longitude and *R* is the radius of the planet at latitude $\varphi$. While the second term generally dominates the vorticity of zonal flows at mid-latitudes, the last term in eqn. 3 cannot be ignored in polar latitudes being comparable to the previous term.

In order to calculate polar two-dimensional maps of vorticity, we have to avoid the singularity of spherical coordinates at the pole. We have done so following the analysis of Venus polar region by *Garate-López et al.* [2013] who introduce a set of local coordinates centered at the pole, $X = (90-\varphi)\cos\lambda$ and $Y = (90-\varphi)\sin\lambda$. These coordinates are local cartesian coordinates where vorticity can be calculated simply by the curl of the two-dimensional velocity field. In our case, we interpolate our measured wind vectors in a regular grid of $X$ and $Y$ coordinates which corresponds to a latitudinal step of 0.3° and after smoothing the result, we calculate the derivatives defining the vorticity as centered finite differences.

[Figure 11]

We show the vorticity maps from this procedure in Figure 11. In panels A and C we plot vorticity for latitudes up to 83 degrees in the north and south respectively. The red area present in both hemispheres shows the anticyclonic side of the jets, while the blue colour shows the cyclonic side. The vorticity of the North jet associated with the hexagon and the eastward jet in the South are similar ~ $6 \pm 1 \times 10^{-5}$ s$^{-1}$. The inhomogeneities observed inside the jet in panel C lie inside the error bars. The red region in panels A and C represent the anticyclone in the North and the SPS in South Polar Region described in section 3.1, with a peak vorticity of ~ $5 \pm 1 \times 10^{-5}$ s$^{-1}$ and ~ $7 \pm 1 \times 10^{-5}$ s$^{-1}$ respectively. Panels B and D in this figure zoom the maps for latitudes > 85° in both hemispheres showing the domain of the intense polar cyclonic vortices with vorticity values dominated by the geometry term and reaching intense values of $-25 \pm 1 \times 10^{-5}$ s$^{-1}$ at 89.8°N.

The relative contributions of the different terms in eqn. 5 to the relative vorticity are represented in Figure 12 in terms of the meridional profiles of relative vorticity. The peak value of $\partial \bar{u}/\partial y$ for the hexagon jet is $5 \pm 1 \times 10^{-5}$ s$^{-1}$, while the eastward jet at 70.4°S reaches a value of $4 \pm 1 \times 10^{-5}$ s$^{-1}$, approximately a tenth of the Coriolis parameter $f = 2\Omega\sin\varphi$ at those latitudes ($\Omega$ is the planetary angular velocity). The contribution of $u \tan \varphi / R$ in eqn. 3 becomes important at latitudes higher than 80º and dominant for latitudes higher than 85º. In the polar vorttices eastward jets, relative vorticity reaches values of the order of magnitude of the Coriolis parameter at the pole, $|f| = 2\Omega = 3.27 \times 10^{-4} \text{s}^{-1}$.

[Figure 12]

5.2 Vorticity gradients

The Rayleigh-Kuo instability criterion for a pure barotropic atmosphere indicates that a zonal jet may become unstable when $\partial^2 \bar{u} / \partial y^2 > \beta$ (a necessary condition though not sufficient), where $\beta = df/dy = 2\Omega \cos \varphi / R(\varphi)$ is the planetary vorticity gradient, with $\Omega$ the planet angular velocity and $R(\varphi)$ is the radius of Saturn at different latitudes [*Sánchez-Lavega,* 2011b].

Figure 13 shows meridional profiles of $\partial^2 \bar{u} / \partial y^2$, a good approximation to meridional vorticity gradients for latitudes lower the 80º. We see in this figure that the $\beta$ parameter, of the order of 1.5x10$^{-12}$ m$^{-1}$s$^{-1}$, is much smaller than the absolute value of average ambient vorticity at the latitudes close to the jets at 70.4ºS and 75.8ºN. Thus, Rayleigh-Kuo criterion for instability is essentially satisfied whenever $\partial^2 \bar{u} / \partial y^2 > 0$, as happens clearly at the two flanks of the eastward jets in both hemispheres.

[Figure 13]

*Barbosa Aguiar et al.* [2010] performed a linear stability analysis of the Rayleigh-Kuo criterion for the hexagon jet on the basis of the terms $\partial^2 \bar{u} / \partial y^2$ and $\beta$ and the Rossby deformation radius $L_D$ [*Sánchez-Lavega,* 2011b]. Their analysis shows how the growth rate of barotropic instabilities depends on zonal wavenumber (*m*) and $L_D$. Since the

North and South jets at the hexagon latitude are similar in shape, the analysis should also hold for the South jet at 70.4°S although we note the difference of 5° in latitude location and the lower peak velocity of Southern jet. According to this simple model, the non-detection (or non-existence) of a wave at the Southern jet at 70.4ºS could be due to the excitation of a wave with a higher wavenumber than 6 and lower amplitude. For example $m > 9$ requires $L_D > 3,000$ km which is reasonable for Saturn polar regions. We leave a further comparative study of stability conditions at both quasi-symmetric jets for a future work.

## 6. Turbulent wind components

The transport of momentum from eddies in a turbulent flow to the zonal jets of giant planets has been the object of intense debate in the literature [*Beebe et al.,* 1980; *Ingersoll et al.,* 1981; *Sromovsky et al.,* 1981; *Salyk et al.,* 2006; *Del Genio et al.,* 2007; *Del Genio and Barbara,* 2012]. If this transport does exist, it implies a positive correlation between northward flux of eastward eddy momentum and the meridional gradient of the zonal flow [*Andrews and McIntyre*, 1976; *Sánchez-Lavega,* 2011b].

In a zonal jet, the average northward flux of eastward momentum is given by the zonal average of $\langle u'v' \rangle$, where $u'_i = u_i - \bar{u}$ and $v'_i = v_i - \bar{v}$, where $u'$ and $v'$ are the zonal and meridional eddy components, $u_i$ and $v_i$ are the components of the wind of the measurement labeled as $i$, and $\bar{u}$ and $\bar{v}$ their zonal averages. We calculate these averages using the same bins as in the case of the mean zonal profile, that is, latitudinal bins of 0.4º every 0.2º in the North hemisphere and bins of 0.6º every 0.3º in the South hemisphere.

In the case of the jet associated with the hexagon we analyze the turbulent eddies at a scale much lower scale than the 6 mode wave. Since the flow is not strictly zonal, $\langle u'v' \rangle$ does not describe adequately the transfer of momentum to the sheared flow. In order to account for this fact, we introduced in section 4 a new coordinate measuring the distance to the maximum of the jet at each longitude, and calculated velocity averages $\bar{u}_h$ and $\bar{v}_h$ using bins of constant value of this new coordinate. The injection of momentum due to turbulent motion can then be estimated as $\langle u'_h v'_h \rangle$, where $u'_{hi} = u_{hi} - \bar{u}_h$ and $v'_{hi} = v_{hi} - \bar{v}_h$, where $u_{hi}$ and $v_{hi}$ are the components tangent and perpendicular to the mean flow of an individual velocity measurement [*Starr*, 1968].

We have analyzed $\langle u'v' \rangle$ in four regions, those with velocity higher than 40 ms$^{-1}$ in the eastward jets at 70.4 ° S and 75.8 ° N and those with velocities lower than 40 ms$^{-1}$ in the westward jets to the north of them. In order to estimate the errors in the zonal averages of $\langle u'v' \rangle$ we have followed *Ingersoll et al.* [1981]:

$$\left(\Delta \langle u'v' \rangle\right)^2 = \frac{1}{n}\left(\sigma_u^{\,2}(\Delta_n u)^2 + \sigma_v^{\,2}(\Delta_n v)^2 + (\Delta_n u)^2 (\Delta_n v)^2\right), \qquad (4)$$

where *n* is the number of measurements in the bin, $\sigma_u$ and $\sigma_v$ are the standard deviations of zonal and meridional velocities and $\Delta u$, $\Delta v$ the basic errors due to navigation and pixel resolution. We obtain errors of the order of 7 m$^2$s$^{-2}$ in all regions analyzed.

**[Figure 14]**

The power per unit mass transmitted to the jet by eddies is given by $\langle u'v' \rangle \partial \bar{u} / \partial y$ [*Andrews and McIntyre*, 1976]. We show in figure 14 scatter plots of $\langle u'v' \rangle$ and $\partial \bar{u} / \partial y$ for the four jets. These figures show a weak correlation between these parameters in the westward jets, and a much smaller correlation in the eastward jets. It seems therefore that the eastward jet embedded in the hexagonal wave and its counterpart jet in the South, are not being fed by eddy momentum injection. The absence of momentum injection by eddies might be caused by injection of energy at smaller scales not resolvable by the measurements or might reinforce the idea that the jets are in fact deeply rooted structures driven by the internal convection [*Aurnou and Olson,* 2001; *Kaspi et al.* 2009] making them insensitive at tropospheric level to the seasonal cycle [*Sánchez-Lavega et al.*, 2014].

## 7. Discussion

In this paper, we have measured and analyzed the horizontal velocity field at cloud level of Saturn polar region from latitudes ~ 60º to 90°. There are obvious analogies between the two hemispheres, such as the presence of a fast polar jet, reaching velocities as high as 140±15 ms$^{-1}$ at latitudes 88.5º and constituting an intense polar vortex (accordingly with *Fletcher et al.* [2008]) of approximately 3,000 km diameter, with very high relative vorticity that approaches the value of planetary vorticity. In both hemispheres there is another fast eastward jet of ~100 ms$^{-1}$ [*Sánchez-Lavega et al.*, 2002; *Vasavada et al.*, 2006; *Baines et al.,* 2009; *García-Melendo et al.,* 2011] whose latitude is significantly different between the two hemispheres, an aspect that could be related with the presence of the hexagon in the North. In the South, a lower latitude of the jet implies that the region between the polar vortex and the jet is considerable larger, allowing the polar jet's velocity to decrease less sharply than in the North. In both hemispheres singular isolated anticyclonic vortices are present, and they are well captured and characterized in vorticity maps.

The most conspicuous difference between the two hemispheres is the presence of the hexagonal feature in the north. A seasonal effect for this major difference between the two polar regions can be ruled out since the hexagon has been observed in the north hemisphere every time high-resolution observations have been available even during the long polar winter night [*Baines et al.* 2009; *Sánchez-Lavega et al.,* 2014], while no hexagon has ever been observed in the South polar region, where only evanescent quasi-linear features appear on some high-resolution observations. An analysis of the two-dimensional velocity field has allowed us to quantify the meandering of the jet associated to the hexagon, showing that it is fitted very well by a sinusoidal wave of wave-number 6 and amplitude 0.5º (117km). The fit reproduces very well the overall hexagonal shape and the detailed curvature of the jet. We also show this jet to be faster than its counterpart in the south, and slightly narrower.

We have also deduced mean velocity profiles for the zonal and meridional velocities, confirming the results of *García Melendo et al.* [2011] and extending them in the northern hemisphere to 89.9º in the North. We provide in the Appendix two tables that summarize our results. In the case of the jet associated to the hexagon, we have studied the effect of the meandering motion of the jet, introducing coordinates adapted to the motion of the jet that show it to be slightly faster than suggested by standard zonal averages.

Finally, we have analyzed the Rayleigh-Kuo instability criterion for a barotropic flow in a shallow atmosphere for the lower latitude eastward jets in the north and south, showing that both display potential instability at their flanks, and we have attempted to detect if there is transfer of momentum from the turbulent motion to the jet. Although in this case we would need much larger statistics to obtain an entirely reliable result, we find no definite traces of momentum transfer into the eastward jets.

Our study leaves a number of questions open to further study. As discussed above, the similarities between the polar zones are remarkable, and there is no indication of why a stable hexagonal wave is present in the north while there is no equivalent wave in the south. The only differences are that the jets lie at different latitudes, with the northern jet closer to the pole, and have slightly different width and peak value. From the dynamical point of view both jets present similar characteristics in terms of the barotropic instability criterion. Seasonal effects affect the thermal structure of both polar regions down to the 300 mbar level [*Fletcher et al.* 2014]. However, the persistence and stability of the polar vortex and the hexagonal jet during the polar night traced from VIMS data sensitive to much deeper atmospheric levels [*Baines et al.* 2009] seem to imply that there are no significant differences between both hemispheres that could make the Rossby deformation lower in the northern hemisphere exciting the $m = 6$ wave. The current presence in the south of the SPS, a big anticyclone of ~7,000 km just north of the eastern jet observed at least from April 2008 to January 2009, and similar to the NPS that existed when the hexagon was first detected by Voyager 1 and 2, could be taken as another hint indicating that it was not the NPS that excited the hexagonal wave. Nevertheless one has to be cautious, because a smaller latitudinal amplitude of the oscillation could make the wave undetectable in the South polar region.

**Acknowledgements**

The data for this paper is available at NASA Planetary Data System (PDS) in COISS volumes 2026, 2047, 2050 and 2083 (http://pds-imaging.jpl.nasa.gov/volumes/iss.html). Data supporting the figures and tables could also be requested from Arrate Antuñano, (arrate.antunano@ehu.es). We are thankful to two anonymous referees that helped to improve this paper. We gratefully acknowledge the work of the Cassini ISS team that allowed these data to be obtained. A.A. is supported by a MINECO FPI PhD Studentship. This work was supported by the Spanish project AYA2012-36666 with FEDER support, PRICI-S2009/ESP-1496, Grupos Gobierno Vasco IT-765-13 and by Universidad del País Vasco UPV/EHU through program UFI11/55.

# References

Acton, Ch. H. (1995), Ancillary data services of NASA's Navigation and Ancillary Information Facility, *Planet. Space Sci.*, **44**, 65-70.

Allison, M., D. A. Godfrey, and R. F. Beebe (1990), A wave dynamical interpretation of Saturn's Polar Hexagon, *Science*, **247**, 1061–1063.

Andrews, D.G. and M.E. McIntyre (1976), Planetary Waves in Horizontal and vertical shear: The generalized Eliassen-Palm relation and the mean zonal acceleration, *Journal of the Atmospheric Sciences*, **33**, 2031-2048.

Aurnou, J.M. and P.L. Olson (2001), Strong zonal winds from thermal convection in a rotating spherical shell. *Geophys. Res. Lett.* **28**, 2557-2559.

Baines, K.H., T.W. Momary, L.N. Fletcher, A.P. Showman, M. Roos-Serote, R. H. Brown, B.J. Buratti, R.N. Clark and P.D. Nicholson (2009), Saturn's north polar cyclone and hexagon at depth revealed by Cassini/VIMS, *Planetary and Space Science*, **57**, 1671-1681.

Barbosa Aguiar A. C., P.L. Read, R.D. Wordsworth, T. Salter and Y.H. Yamazaki (2010), A laboratory model of Saturn's North Polar Hexagon, *Icarus*, **206**, 755-763.

Beebe, R.F., A.P. Ingersoll, G.E. Hunt, J.L. Mitchell and J-P. Müller (1980), Measurements of wind vectors, eddy momentum transports and energy conversions in Jupiter´s atmosphere from Voyager 1 images, *Geophysical Research Letters*, **7**,1-4.

Caldwell, J., X-M. Hua, B.Turgeon, J.A. Westphal and C.D. Barnet (1993), The drift of Saturn's North Polar Spot by the Hubble Space Telescope, *Science*, **260**, 326-329.

Del Genio, A. D., J.M. Barbara, J. Ferrier, A. P. Ingersoll, R. A. West, A. R. Vasavada, J. Spitale and C. C. Porco (2007), Saturn eddy momentum fluxes and convection: First estimates from Cassini images, *Icarus*, **189**, 479-492.

Del Genio, A.D., R.K Achterberg, K. H. Baines, F. M. Flasar, P. L. Read, A. Sánchez-Lavega and A. P. Showman (2009), Saturn Atmospheric Structure and Dynamics, in Saturn from Cassini–Huygens, M. K. Dougherty, L. W. Esposisto and S.M. Krimigis, Eds. Springer-Verlag, 113-159, London

Del Genio, A.D. and J.M. Barbara (2012), Constraints on Saturn's tropospheric general circulation from Cassini ISS images, *Icarus*, **219**, 689-700.

Desch M. D., L. M. Kaiser (1981) Voyager measurements of the rotation period of Saturn's magnetic field, *Geophys. Res. Lett.*, **8**, 253-256.

Dyudina, U.A., A. P. Ingersoll, S. P. Ewald, A. R. Vasavada, R. A. West, A. D. Del Genio, J. M. Barbara, C. C. Porco, R. K. Achterberg, F. M. Flasar, A. A. Simon-Miller and L. N. Fletcher (2008), Dynamics of Saturn's South Polar Vortex, *Science* **319**, 1801.

Dyudina, U.A., A.P. Ingersoll, S.P. Ewald, A.R. Vasavada, R.A. West, K.H. Baines, T.W. Momary, A.D. Del Genio, J.M. Barbara, C.C. Porco, R.K. Achterbebrg, F. M. Flasar, A.A. Simon-Miller and L.N. Fletcher (2009), Saturn's south polar vortex compared to other large vortices in the Solar System, *Icarus,* **202**, 240-248.

Dyudina, U.A., A.P. Ingersoll, S.P. Ewald, C.C. Porco, G. Fischer, Y. Yair (2013), Saturn's visible lightning, its radio emissions, and the structure of the 2009-2011 lightning storms, *Icarus*, **226**, 1020-1037.

Emanuel, K.A. (1994), Atmospheric Convection, Oxford University Press, Oxford.

Fletcher, L.N., P.G.J. Irwin, G.S. Orton, N.A. Teanby, R.K. Achterberg, G.L. Bjoraker, P.L. Read, A.A. Simon-Miller, C. Howett, R. de Kok, N. Bowles, S.B. Calcutt, B. Hesman and F.M. Flasar (2008), Temperature and composition of Saturn's polar hot spots and hexagon, *Science*, **319**, 79-81.

Fletcher, L. N., J.A. Sinclair, P.G.J. Irwin, R.S. Giles, G.S. Orton, B.E. Hesman, J. Hurley, G.L. Bjoraker, A.A. Simon (2014), Seasonal Evolution of Saturn's Polar Atmosphere from a Decade of Cassini/CIRS Observations, *Icarus*, in press.

Garate-Lopez, I., R. Hueso, A. Sánchez-Lavega, J. Peralta , G. Piccioni & P. Drossart (2013) A chaotic long-lived vortex at the southern pole of Venus. *Nature Geoscience*, **6**, 254-257

García-Melendo, E., A. Sánchez-Lavega, J.F. Rojas, S. Pérez-Hoyos and R.Hueso (2009), Vertical shears in Saturn's eastward jets at cloud level, *Icarus*, **201**, 818-820.

García-Melendo E., S. Pérez-Hoyos, A. Sánchez-Lavega and R. Hueso (2011), Saturn's zonal wind profile in 2004 - 2009 from Cassini ISS images and its long-term variability, *Icarus,* **215**, 62-74.

García-Melendo E., R. Hueso, A. Sánchez-Lavega, J. Legarreta, T. del Río-Gaztelurrutia, S. Pérez-Hoyos and J. F. Sanz-Requena (2013), Atmospheric dynamics of Saturn's 2010 giant storm", *Nature Geoscience,* **6**, 525–529.

Godfrey, D. A. (1988), A hexagonal feature around Saturn's north pole, *Icarus*, **76**, 335–356.

Gurnett D. A. J. B. Groene, A. M. Persoon, J. D. Menietti, S. Y. Ye, W. S. Kurth, R. J. MacDowall and A. Lecacheux (2010), The reversal of the rotational modulation rates of the north and south components of Saturn kilometric radiation near equinox, *Geophys. Res. Lett.*, **37**, L24101.

Houze R.A. Jr (1993), Cloud Dynamics, Academic Press, INC., A division of Harcourt Brace & Company, San Diego, California.

Hueso, R., J. Legarreta, E. García-Melendo, A. Sánchez-Lavega, and S. Pérez-Hoyos (2009), The Jovian anticyclone BA: II. Circulation and models of its interaction with the zonal jets", *Icarus*, **203**, 499–515.

Hueso, R., J. Legarreta, J. F. Rojas, J. Peralta, S. Pérez-Hoyo, T. del Río-Gaztelurrutia, and A. Sánchez-Lavega (2010), The Planetary Laboratory for Image Analysis (PLIA), *Adv. Space Res.*, **6**, 1120–1138.

Hunt G. E. and P. Moore (1982), "Saturn", Rand-McNally, London

Ingersoll, A.P., R.F. Beebe, J.L. Mitchell, G.W. Ganeau, G.M. Yagi and J-P. Müller (1981), Interaction of Eddies and Mean Zonal Flow on Jupiter as Inferred from Voyager 1 and 2 images, *Journal of Geophysical Research*, **86**, A10, 8733-8743.

Kaspi, Y.; G. R. Flierl, A. P. Showman (2009), The deep wind structure of the giant planets: Results from an anelastic general circulation model, *Icarus*, **202**, 525-542.

Limaye, S.S., H.E. Revercomb, L.A. Sromovsky, R.J. Krauss, D.A. Santek and V.E. Suomi (1982), Jovian Winds from Voyager 2: Part I: Zonal Mean Circulation, *Journal of the Atmospheric Sciences*, **39**, 1413-1432.

Luz, D., D. L. Berry, G. Piccioni, P. Drossart, R. Politi, C. F. Wilson, S. Erard and F. Nuccilli (2011), Venus's southern polar vortex reveals precessing circulation. *Science* **332**, 577–580.

Mitchell, D. M., L. Montabone, S. Thomson, and P.L. Read (2014), Polar vortices on Earth and Mars: A comparative study of the climatology and variability from reanalyses. *Q.J.R. Meteorol. Soc.*. doi: 10.1002/qj.2376

Orton, G.S. and P.A. Yanamandra-Fisher (2005), Saturn's Temperature Field from High-Resolution Middle-Infrared Imaging, *Science*, **307**, 696-698.

Pérez-Hoyos, S., A. Sánchez-Lavega (2006), Solar flux in Saturn's atmosphere: maximum penetration and heating rates in the aerosol and cloud layers, *Icarus*, **180**, 368-378 (2006).

Pérez-Hoyos S., A. Sánchez-Lavega, R. G. French, and J. F. Rojas (2005), Saturn's cloud structure and temporal evolution from ten years of Hubble Space Telescope Images (1994-2003), *Icarus*, **176**, 155-174.

Piccioni, G., P. Drossart, A. Sanchez-Lavega, R. Hueso, F. W. Taylor, C. F. Wilson, D. Grassi, L. Zasova, M. Moriconi, A. Adriani, S. Lebonnois, A. Coradini, B. Bézard, F. Angrilli, G. Arnold, K. H. Baines, G. Bellucci, J. Benkhoff, J. P. Bibring, A. Blanco, M. I. Blecka, R. W. Carlson, A. Di Lellis, T. Encrenaz, S. Erard, S. Fonti, V. Formisano, T. Fouchet, R. Garcia, R. Haus, J. Helbert, N. I. Ignatiev, P. G. J. Irwin, Y. Langevin, M. A. Lopez-Valverde, D. Luz, L. Marinangeli, V. Orofino, A. V. Rodin, M. C. Roos-Serote, B. Saggin, D. M. Stam, D. Titov, G. Visconti, M. Zambelli1 & the VIRTIS-Venus Express Technical Team (2007), South-polar features on Venus similar to those near the north pole. *Nature*, **450**, 637–40.

Porco, C.C., R.A. West, S. Squyres, A. McEwen, P. Thomas, C.D. Murray, A. Del Genio, A.P. Ingersoll, T.V. Johnson, G. Neukum, J. Veverka, L. Dones, A. Brahic, J.A. Burns, V. Haemmerle, B. Knowles, D. Dawson, T. Roatsch, K. Beurle, and W. Owen

(2004), Cassini imaging science: Instrument characteristics and anticipated scientific investigations at Saturn, *Space Sci. Rev.*, **115**, 363–497.

Porco, C.C., E. Baker, J. Barbara, K. Beurle, A. Brahic, J.A. Burns, S. Charnoz, N. Cooper, D.D. Dawson, A.D. Del Genio, T. Denk, L. Dones, U. Dyudina, M.W. Evans, B. Giese, K. Grazier, P. Helfenstein, A.P. Ingersoll, R.A. Jacobson, T.V. Johnson, A. McEwen, C.D. Murray, G. Neukum, W.M. Owen, J. Perry, T. Roatsch, J. Spitale, S. Squyres, P. Thomas, M. Tiscareno, E. Turtle, A.R. Vasavada, J. Veverka, R. Wagner, and R. West (2005), Cassini Imaging Science: Initial Results on Saturn's Atmosphere. *Science*, **307**, 1243-1247.

Salyk, C., A.P. Ingersoll, J. Lorre, A. Vasavada. and A.D. Del Genio (2006), Interaction between eddies and mean flow in Jupiter's atmosphere: Analysis of Cassini imaging data, *Icarus*, **185**, 430-442.

Sánchez-Lavega, A., J. Lecacheux, F. Colas and P. Laques (1993), Ground-Based observations of Saturn's North Polar Spot and Hexagon, *Science*, **260**, 329-332.

Sánchez Lavega A., J. F. Rojas, J. R. Acarreta, J. Lecacheux, F. Colas and P. V. Sada (1997), New Observations and studies of Saturn's long-lived North Polar Spot, *Icarus*, **128**, 322-334.

Sánchez-Lavega, A., J. F. Rojas and P. V. Sada (2000), Saturn's zonal winds at cloud level, *Icarus*, **147**, 405-420.

Sánchez-Lavega, A., S. Pérez-Hoyos, J.R. Acarreta, R. G. French (2002), No hexagonal wave around Saturn's Southern Pole, *Icarus*, **160**, 216-219.

Sánchez-Lavega, A., Pérez-Hoyos, S., Rojas, J. F., Hueso, R., French, R.G (2003), A strong decrease in Saturn's equatorial jet at cloud level, *Nature*, **423**, 623-625.

Sánchez-Lavega, A., R. Hueso, S. Pérez-Hoyos, J. F. Rojas, and R. G. French (2004), Saturn's cloud morphology and zonal winds before the Cassini encounter. *Icarus*, **170**, 519-523.

Sánchez-Lavega A. (2005), How long is the day on Saturn? *Science*, **307**, 1223-1224.

Sánchez-Lavega, A., R. Hueso, S. Pérez-Hoyos and J. F. Rojas (2006), A strong vortex in Saturn's south pole, *Icarus*, **184**, 524-531.

Sánchez-Lavega, A., T. del Río-Gaztelurrutia, R. Hueso, J. M. Gómez-Forrellad, J. F. Sanz-Requena, J. Legarreta, E. García-Melendo, F. Colas, J. Lecacheux, L. N. Fletcher, D. Barrado-Navascués, D. Parker and the International Outer Planet Watch Team (2011a), Deep winds beneath Saturn's upper clouds from a seasonal long-lived planetary-scale storm, *Nature*, **475**, 71-74.

Sánchez-Lavega, A. (2011b), An Introduction to Planetary Atmospheres, Taylor & Francis, CRC Press, Boca Raton, Florida.

Sánchez-Lavega, A., T. del Río-Gaztelurrutia, R. Hueso, S. Pérez-Hoyos, E. García-Melendo, A. Antuñano, I. Mendikoa, J. F. Rojas, J. Lillo, D. Barrado-Navascués, J. M. Gomez-Forrellad, C. Go, D. Peach, T. Barry, D. P. Milika, P. Nicholas and A. Wesley (2014), The long-term steady motion of Saturn's hexagon amd the stability of its enclosed jet stream under seasonal changes, *Geophysical Reseach Letters*, **41**, 1425-1431.

Sayanagi K M, U.S. Dyudina, S.P. Ewald, G. Fischer, A.P. Ingersoll, W.S. Kurth, G.D. Muro, C.C. Porco, R.A. West (2013a), Dynamics of Saturn's great storm of 2010–2011 from Cassini ISS and RPWS, *Icarus*, **223**, 460-478.

Sayanagi, K.M. S. P. Ewald, U. A. Dyudina and A. P. Ingersoll (2013b). American Astronomical Society, DPS meeting #45, #509.06. 10.

Sayanagi, K.M., U.A. Dyudina, S.P. Ewald, G.D. Muro, and A.P. Ingersoll (2014), Cassini ISS Observation of Saturn's String of Pearls, *Icarus*, **229**, 170-180.

Seidelmann, P. K., B. A. Archinal, M. F. A'Hearn, A. Conrad, G. J. Consolmagno, D. Hestroffer, J. L. Hilton, G. A. Krasinsky, G. Neumann, J. Oberst, P. Stooke, E. F. Tedesco, D. J. Tholen, P. C. Thomas and I. P. Williams (2007), Report of the IAU/IAG working group on cartographic coordinates and rotational elements: 2006, *Celestial Mech. Dyn. Astron.* **98**, 155-180.

Snyder, J. P. (1987) Map Projections-A Working Manual, U. S. Geological Survey Professional Paper 1395. Washington, DC: U. S. Government Printing Office, 191-202.

Sromovsky, L.A., H.E. Revercomb, V.E. Suomi, S.S. Limaye and R.J. Krauss (1981), Jovian Winds from Voyager 2. Part II: Analysis of Eddy Transports, *Journal of the Atmospheric Sciences*, **39**, 1433-1445.

Starr, V.P. (1968) Physics of negative viscosity (with applications to earth and solar atmosphere, spiral galaxies and oceanic circulations), McGraw-Hill.

Teanby, N., P. Irwin, Rd. Kok, & C. Nixon. (2010), Seasonal changes in Titan's polar trace gas abundance observed by Cassini. *Astrophysical Journal Letters*, **724**, L84 - L89.

Trammell, H.J., L. Li, X. Jiang, M. Smith, S.Horst and A. Vasavada (2014), The Global Vortex Analysis of Jupiter and Saturn Based on Cassini Imaging Science Subsystem, Icarus; DOI: 10.1016/j.icarus.2014.07.019

Vasavada, A.W., S.M. Hörst, M.R. Kennedy, A.P. Ingersoll, C.C. Porco, A.D. Del Genio, R.A. West (2006), Cassini imaging of Saturn: Southern hemisphere winds and vortices, *Journal of Geophysical Research*, **111**, E05004.

**Figures & Figure captions**

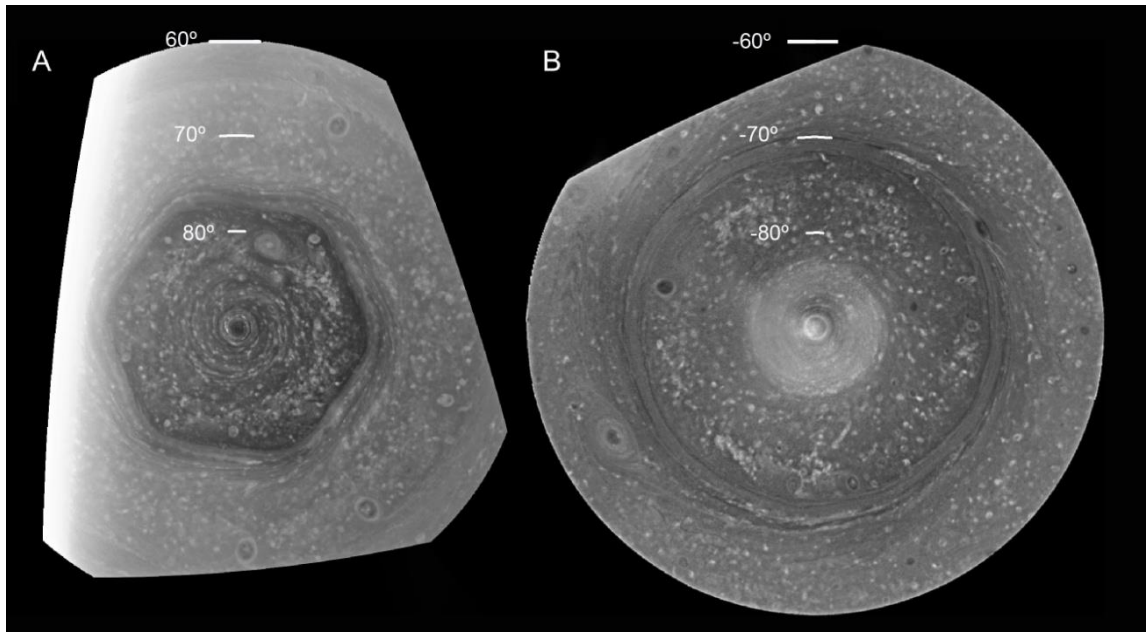

**Figure 1.** Panel A**:** Polar projection of the North Polar Region from 60º to 90º from a Wide Angle Camera image obtained with a CB2 filter on 14 June 2013. Panel B: An equivalent projection of the South Polar Region, from -60º to -90º built using four different images captured by the Wide Angle Camera with a CB2 filter on 3 December 2008. Geometry and illumination conditions are summarized in Table 1.

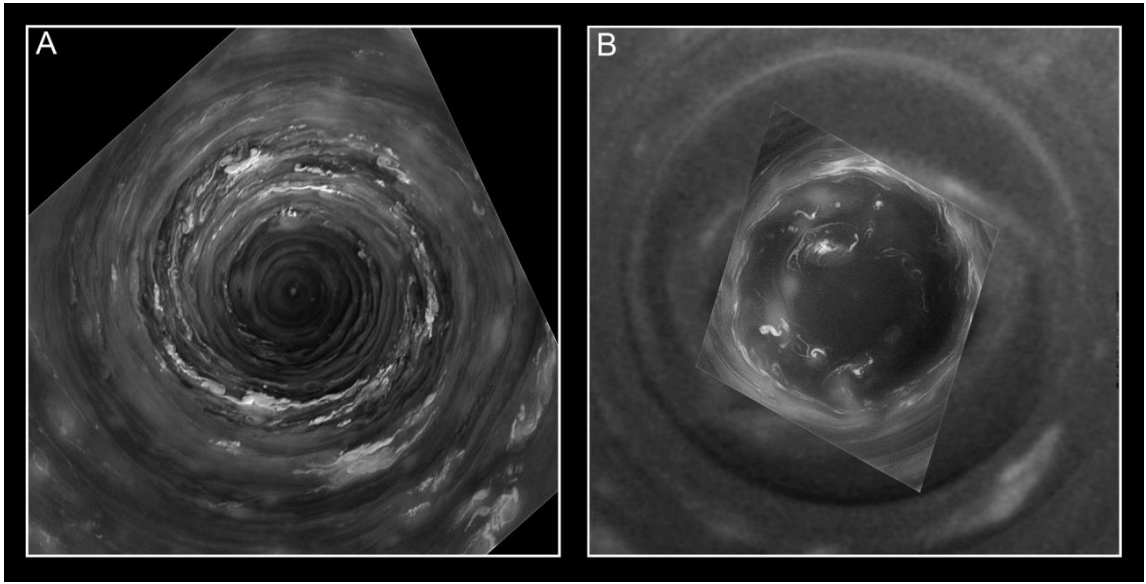

**Figure 2.** Panel A: Polar projection of the North Polar Region from 87.5º to 90º on 14 June 2013, using an image with CB2 filter and spatial resolution at the Pole of 5.3 km/pixel. Panel B: Polar projection of the South Polar Region from 87.5º to 90º on 14 July 2008. This figure in panel B is a composition of two different images with different resolutions. The image on the front is a NAC image with CB2 filter and a spatial resolution of 2.4 km/pixel while the one in the back is a WAC image with a CB2 filter with a spatial resolution of 29.5 km/pixel. Geometry and illumination conditions are summarized in Table 1.

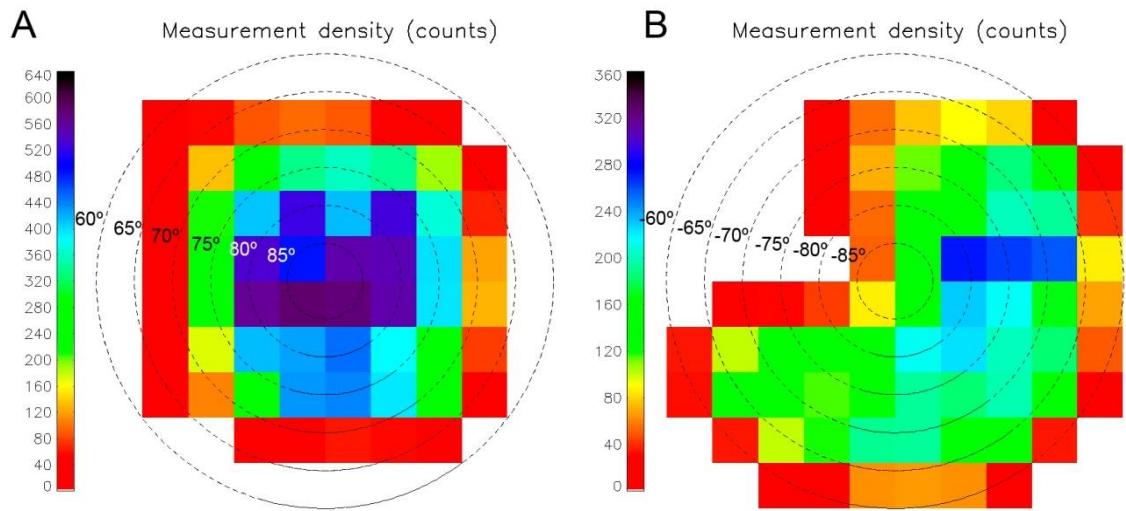

**Figure 3**: Measurement density of wind vectors, displayed as number of counts in square bins of 6 degrees in latitude. Panels A and B depict the north (June 2013) and south (October 2008 and July 2006) polar regions respectively.

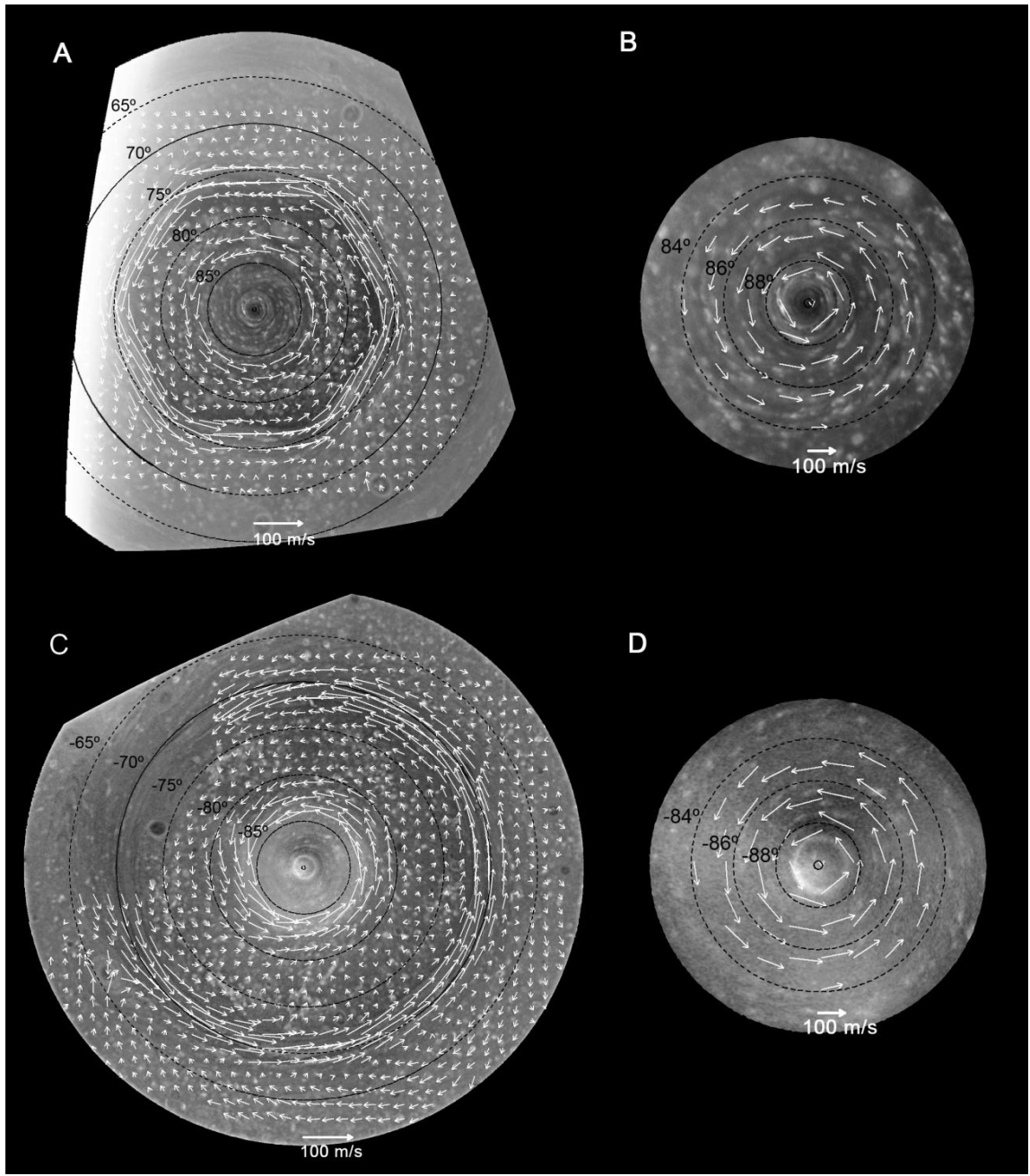

**Figure 4:** Average velocity field for all the dates analyzed. A) North Polar Region from 60ºN to 83ºN; B) North Polar Region from 83º to 90ºN; C) South Polar Region from -60ºS to -90ºS; D) South Polar Region from -83ºS to -90ºS. Note that the vector scales in panels B and D is different to the scale in panels A and C.

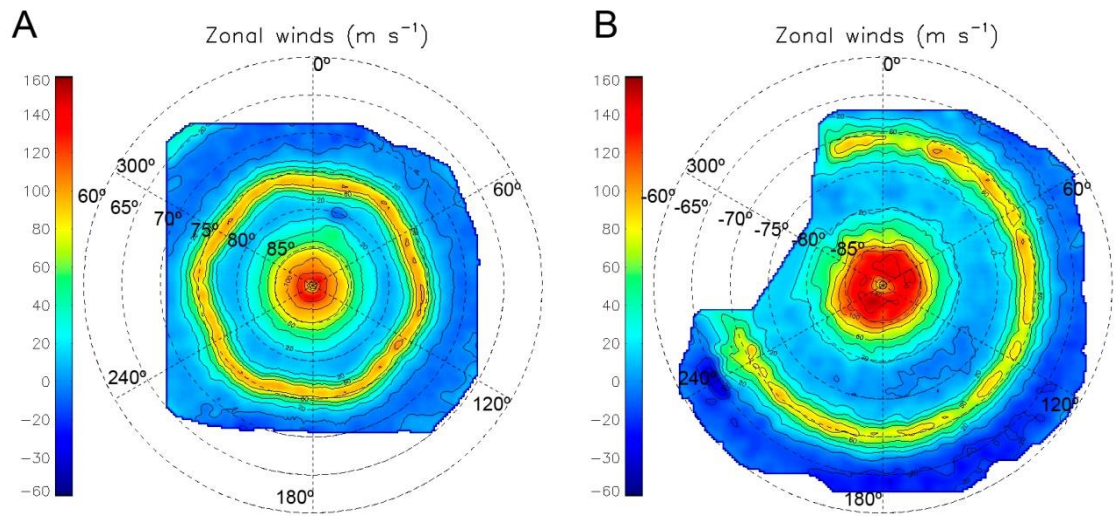

**Figure 5:** Polar maps of zonal wind velocities averaged over all epoch of observation. Panel A: North Polar Region. Panel B: South Polar Region. Individual measurement errors are estimated as 5-10 ms$^{-1}$, which agrees with the statistical analysis of wind dispersions at a given latitude except for the jet cores where winds dispersions are of the order of 15-20 ms$^{-1}$.

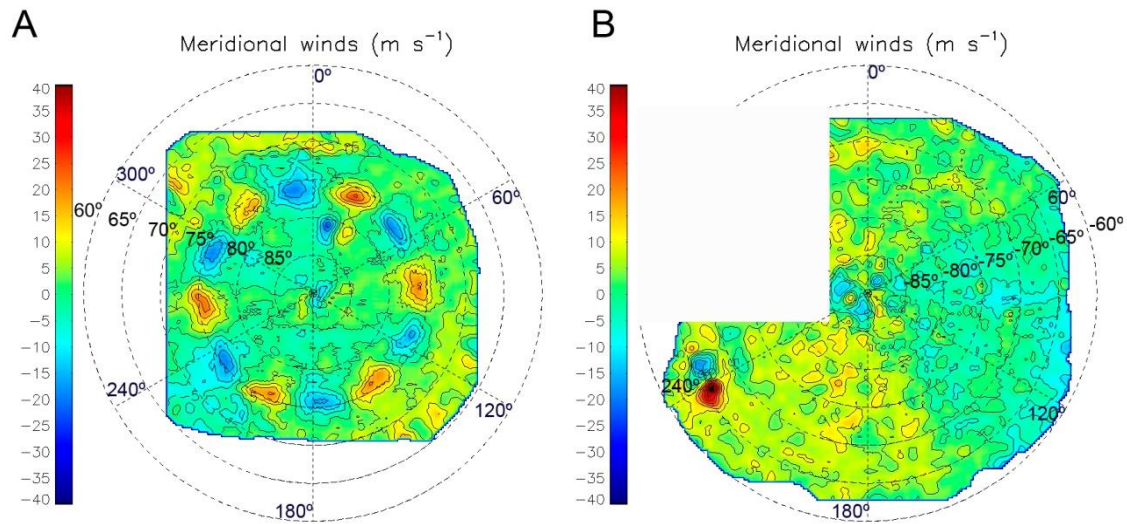

**Figure 6:** Polar maps of meridional wind velocities averaged over all epochs of observations. Panel A: North Polar Region. Panel B: South Polar Region. Positive values are for northward motions. Individual measurement errors are estimated as 5-10 ms$^{-1}$, which also corresponds to the dispersion of wind velocities at all latitudes except for the hexagon. The north vortex at 80º and the SPS at 66ºS have been measured on image pairs from only one date (14 June 2013 for the north vortex and 3 December 2008 for the SPS) and their mean circulation appears only once in this map. The mean global structure visible on Panel B is an artifact of the data at the level of the measurement error and is not significant.

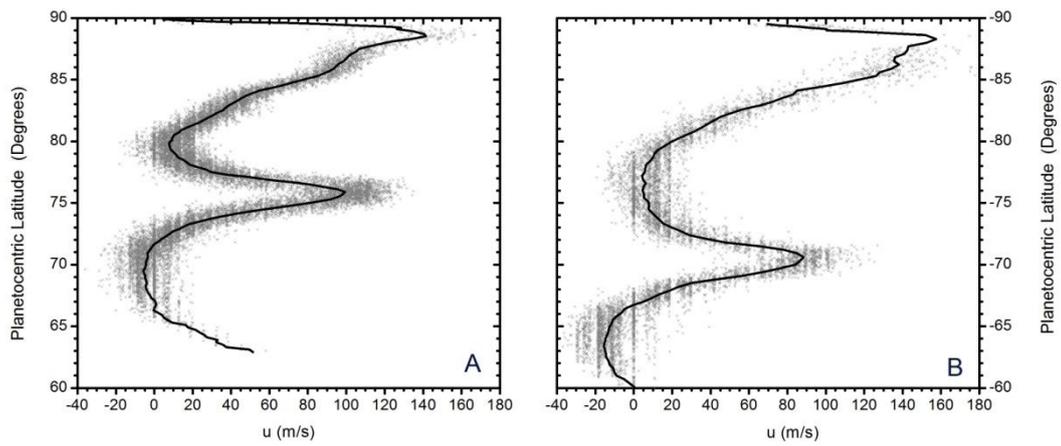

**Figure 7:** Zonal wind profiles of the North Polar Region (A) and South Polar Region (B). In both figures grey points represent the measured zonal vectors while the black line represents the mean zonal velocity averaged over all data in each bin.

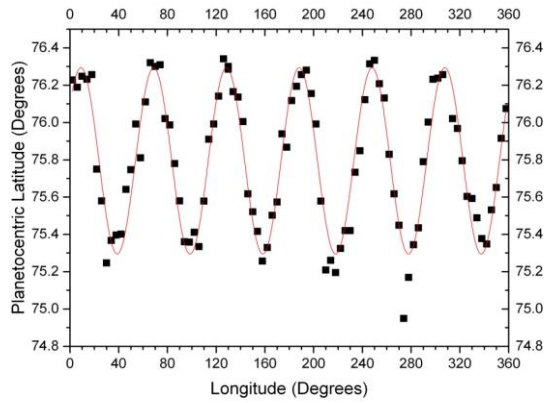 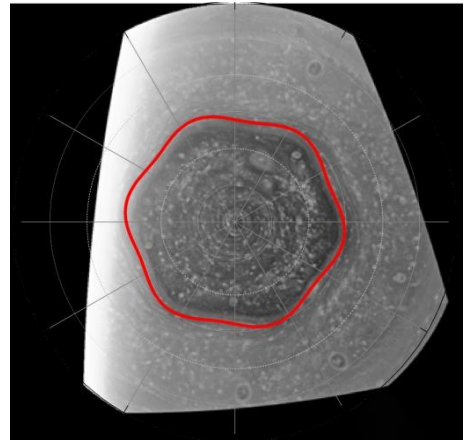

**Figure 8.** A: Deviation of the latitudes of velocity module peaks for different longitudes. The line represents a sinusoidal fit. B: The wave represented by equation (2) plotted in polar coordinates with the background of the polar projection of the north region from 25 December 2013. The dots in panel A at longitude 280º that depart from the sinusoidal fit correspond to the low illuminated area on the leftmost hexagon vertex and represent the effect of noise in that longitude.

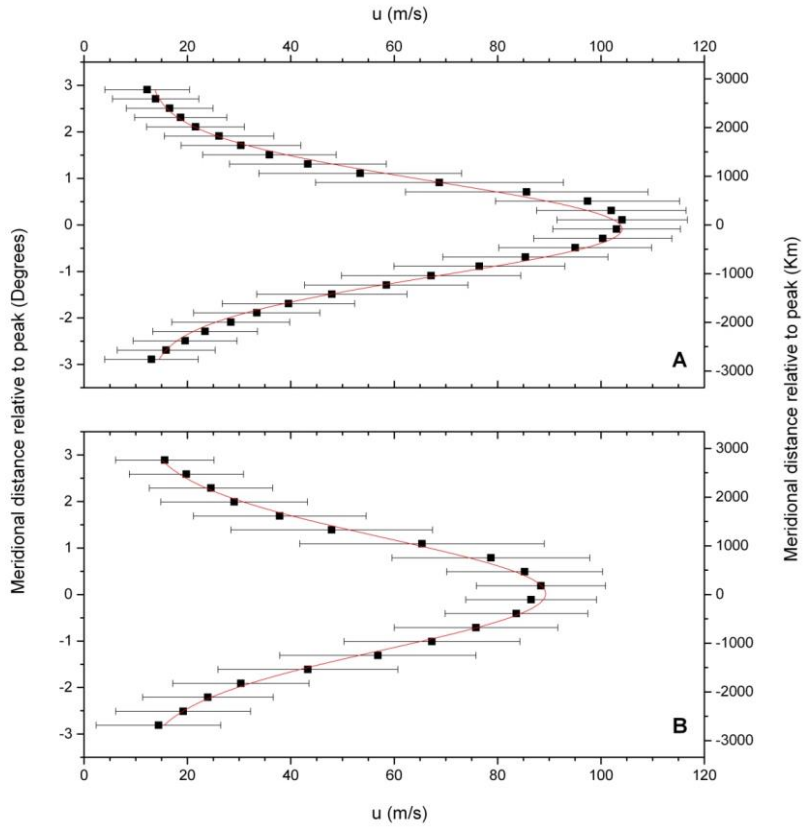

**Figure 9.** Measured winds of eastward jets (black dots) and Gaussian fits (red solid line, see text). Upper panel: Jet at 75.79°± 0.01°N, with a FWHM of 2.38°±0.05° (2,275±50 km), $u_0 = 104 \pm 12$ ms$^{-1}$, and $b = 8.0 \times 10^{-11}$ m$^{-1}$s$^{-1}$. Lower panel: Jet at 70.41° ± 0.01°S, with a FWHM of 2.8° ± 0.1° (2,700±90 km), $u_0 = 88 \pm 20$ ms$^{-1}$ and $b = 5.0 \times 10^{-11}$ m$^{-1}$s$^{-1}$.

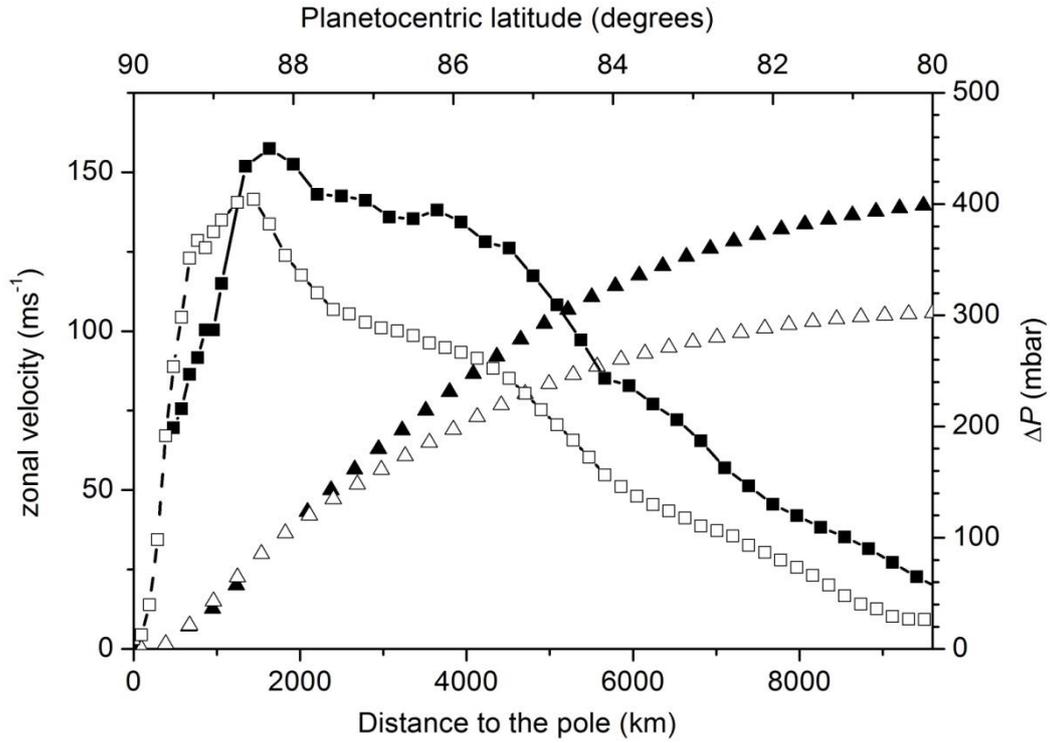

**Figure 10.** A detail of the zonal wind profile at very high latitudes. Black squares correspond to the Southern hemisphere, and white squares to the north. Triangles denote $\Delta P = P(\varphi) - P_{\text{pole}}$, as deduced from the gradient wind balance, with black triangles corresponding to the south and white triangles to the north.

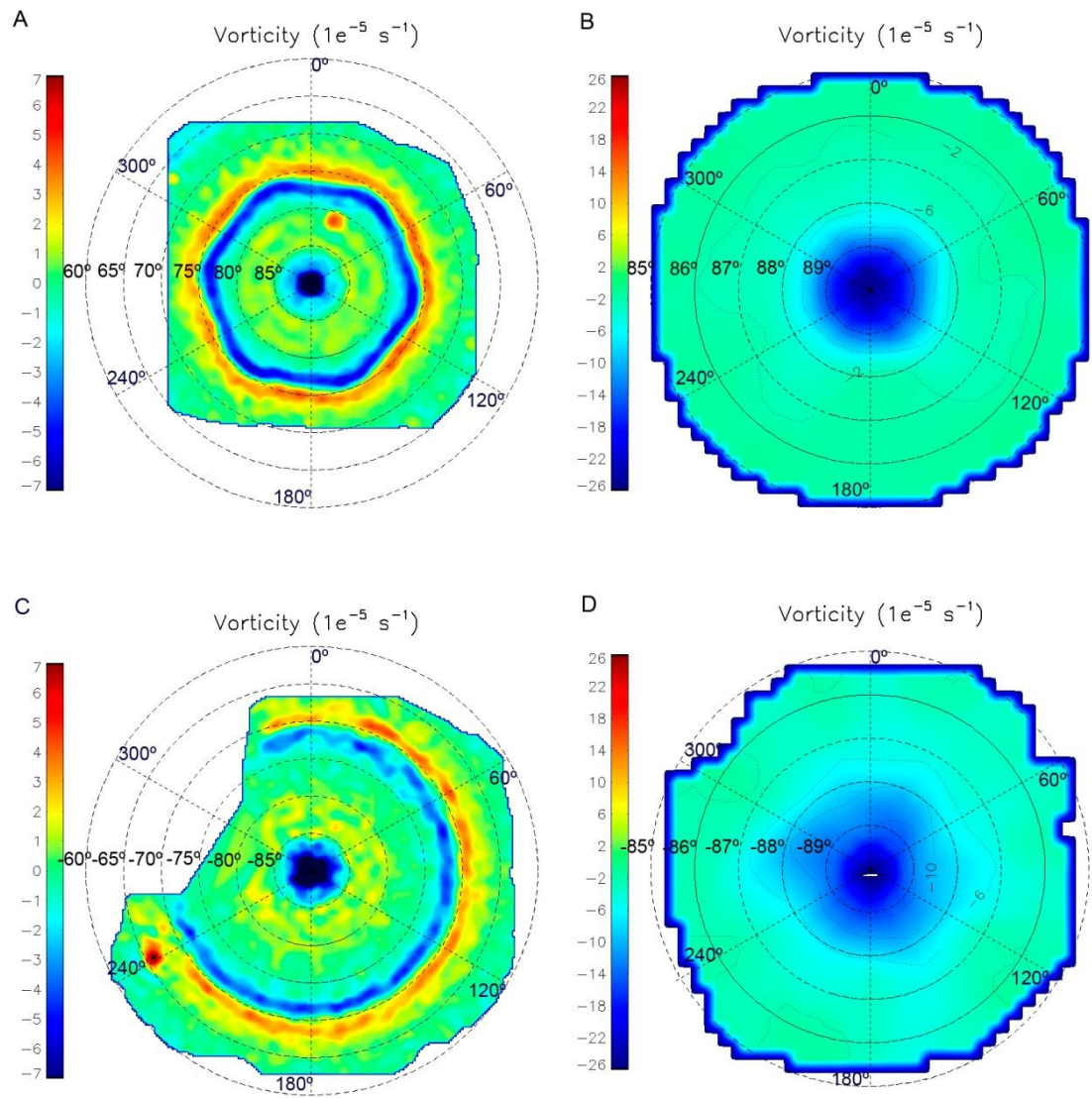

**Figure 11**. Vorticity maps of the polar regions. A) North polar region from 60ºN to 90ºN, B) North Polar Region from 85ºN to 90ºN, C) Soth Polar Region from 60ºS to 90ºS and D) South Polar Region from 83ºS to 90ºS. Blue represents cyclonic motion and red anticyclonic motion. Note that color scales are different for panels A and C compared with B and D. The estimated error is $1 \times 10^{-5}$ s$^{-1}$ for both hemispheres.

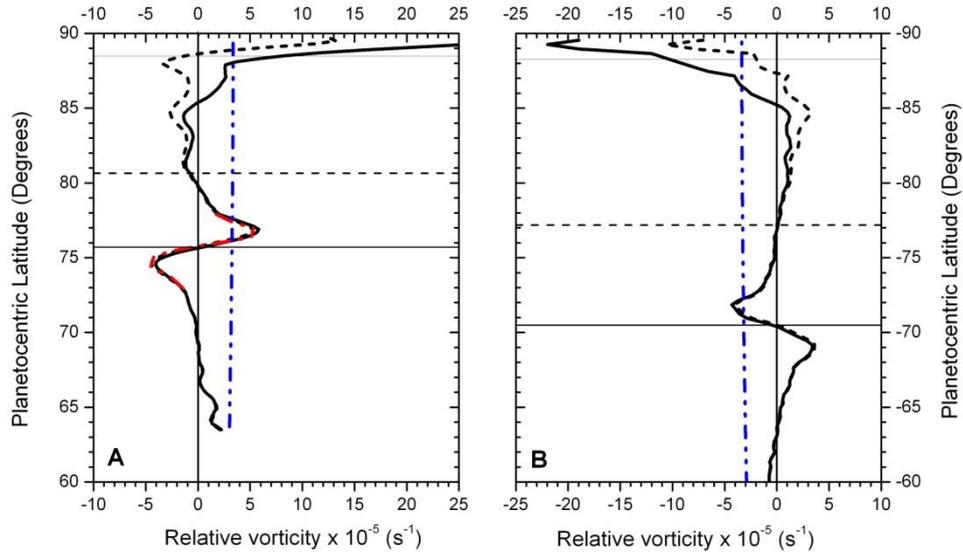

**Figure 12.** Meridional profiles of relative vorticity (solid lines), $-\partial \bar{u}/\partial y$ (dashed line) and Coriolis parameter $f = 2\Omega \sin\varphi$ divided by ten (blue dash-dot line) for the north (panel A) and south (panel B) polar regions. Differences between the relative vorticity and the meridional shear of the zonal wind (dashed line) are due to the geometric term in the vorticity equation. Horizontal solid and dashed lines represent the maxima of eastward and westward jets respectively. Red dashed line in panel A indicates the derivative $-\partial \bar{u}_h/\partial y$, that is, the meridional gradient of the hexagon zonal jet velocity corrected to take into account the hexagon meandering effect.

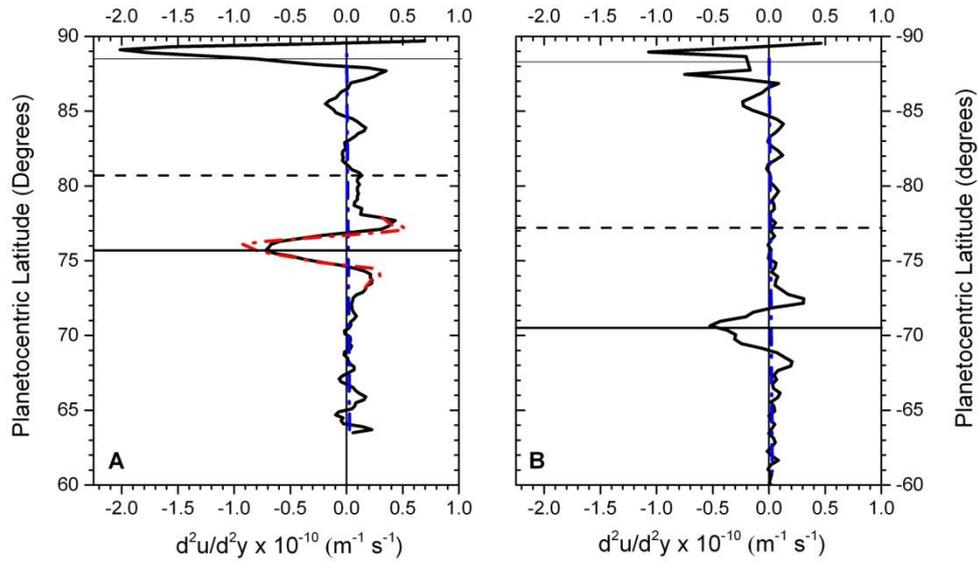

**Figure 13.** A) Vorticity gradient of the North Polar Region and B) vorticity gradient of the South Polar Region represented by the black solid line. The blue dashed line represents the $\beta$ parameter. The dashed horizontal line represents the westward jet and the solid horizontal lines represent the eastward jets. Red dashed line in panel A indicates the $-\partial^2 \bar{u}_h / \partial y^2$ term, that is, the meridional gradient of vorticity of the hexagon jet corrected to take into account the hexagon meandering effect.

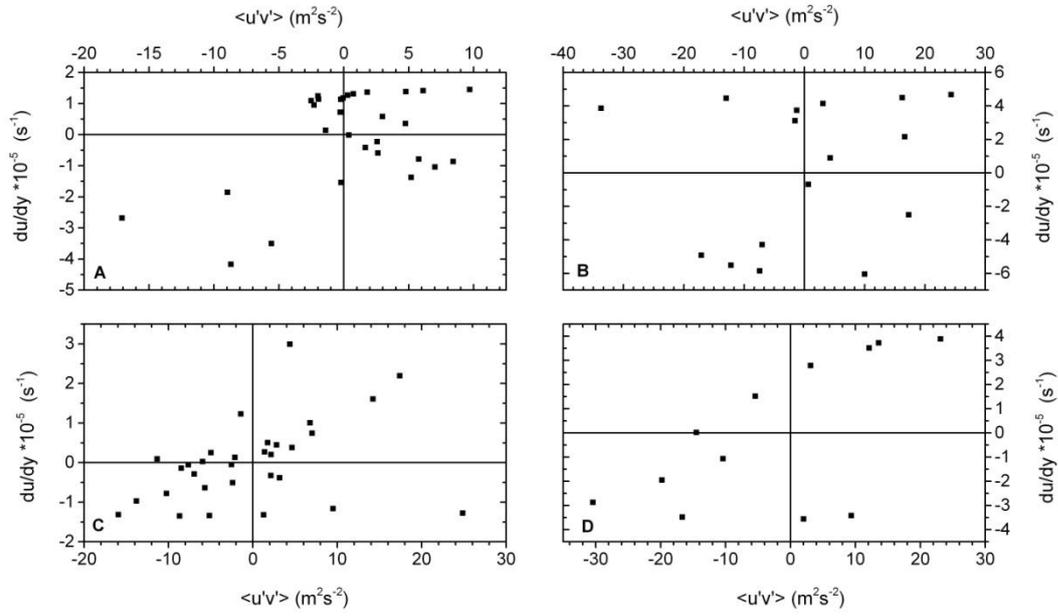

**Figure 14:** Scatter plots of $\langle u'v' \rangle$ and $\partial \overline{u}/\partial y$. A) North westward jet (latitudes 77°-83°); B) North eastward jet (latitudes 74°-77°); C) South westward jet (latitudes 72-81°); D) South eastward jet (latitudes 69°-72°).

**Tables**

**Table 1.** Images used in this study.

| COISS | Image ID | Date | Time | Camera | Filter | Resolution (sub-S/C) (km/pixel) | Resolution at the pole (km/pixel) | Sub-Solar Latitude (º) | Sub-S/C Latitude (º) | Phase angle (º) | Region Studied |
|---|---|---|---|---|---|---|---|---|---|---|---|
| 2026 | W1539288428_1 | 11/10/2006 | 19:35:10 | WAC | CB2 | 17.5 | 25.7 | -16 | 43 | 37 | South |
| 2026 | W1539290777_1 | 11/10/2006 | 20:14:19 | WAC | CB2 | 17.1 | 26.1 | -16 | 41 | 32 | South |
| 2026 | W1539296593_1 | 11/10/2006 | 21:51:15 | WAC | CB2 | 16.5 | 29.5 | -16 | 34 | 20 | South |
| 2026 | W1539297244_1 | 11/10/2006 | 22:02:06 | WAC | CB2 | 16.5 | 29.5 | -16 | 34 | 19 | South |
| 2047 | N1594753107_1 | 14/07/2008 | 18:20:20 | NAC | CB2 | 1.80 | 2.06 | -06 | 61 | 65 | South |
| 2047 | N1594756809_1 | 14/07/2008 | 19:22:02 | NAC | RED | 1.99 | 2.40 | -06 | 56 | 60 | South |
| 2050 | W1607021323_1 | 03/12/2008 | 18:09:09 | WAC | CB2 | 45.3 | 77.1 | -04 | 36 | 45 | South |
| 2050 | W1607026963_1 | 03/12/2008 | 19:43:09 | WAC | CB2 | 46.5 | 83.2 | -04 | 34 | 45 | South |
| 2050 | W1607032604_1 | 03/12/2008 | 21:17:10 | WAC | CB2 | 47.7 | 90.0 | -04 | 32 | 44 | South |
| 2050 | W1607038244_1 | 03/12/2008 | 22:51:10 | WAC | CB2 | 49.0 | 98.0 | -04 | 30 | 42 | South |
| 2050 | W1607043884_1 | 04/12/2008 | 00:25:10 | WAC | CB2 | 50.1 | 103 | -04 | 29 | 41 | South |
| 2083 | N1749893515_1 | 14/06/2013 | 08:36:26 | NAC | CB2 | 4.21 | 5.27 | 19 | 53 | 46 | North |
| 2083 | N1749898322_1 | 14/06/2013 | 09:56:33 | NAC | CB3 | 8.57 | 11.0 | 19 | 51 | 43 | North |
| 2083 | N1749906144_1 | 14/06/2013 | 12:06:55 | NAC | CB2 | 4.40 | 5.83 | 19 | 49 | 39 | North |
| 2083 | N1749908590_1 | 14/06/2013 | 12:47:41 | NAC | CB3 | 8.87 | 11.9 | 19 | 48 | 38 | North |
| 2083 | W1749893515_1 | 14/06/2013 | 08:36:26 | WAC | CB2 | 42.1 | 52.7 | 19 | 53 | 46 | North |
| 2083 | W1749898649_1 | 14/06/2013 | 10:02:00 | WAC | CB2 | 42.8 | 55.1 | 19 | 51 | 43 | North |
| 2083 | W1749916412_1 | 14/06/2013 | 14:58:03 | WAC | CB2 | 45.4 | 64.2 | 19 | 45 | 34 | North |
| 2083 | W1749921546_1 | 14/06/2013 | 16:23:37 | WAC | CB2 | 46.2 | 66.5 | 19 | 44 | 31 | North |
| 2083 | W1750886837_1 | 25/06/2013 | 20:31:39 | WAC | CB3 | 73.7 | 86.0 | 19 | 59 | 72 | North |
| 2083 | W1750894760_1 | 25/06/2013 | 22:43:45 | WAC | CB2 | 37.7 | 44.0 | 19 | 59 | 66 | North |
| 2083 | W1750894808_1 | 25/06/2013 | 22:44:33 | WAC | CB3 | 75.5 | 88.1 | 19 | 59 | 66 | North |
| 2083 | W1750902779_1 | 26/06/2013 | 00:57:21 | WAC | CB3 | 77.5 | 91.4 | 19 | 58 | 61 | North |

**Table 2.** Error for winds tracked with the correlation algorithm.

| Date | Region | Image type | Single pix error (ms$^{-1}$) | Navigation error (ms$^{-1}$) | Estimated error (ms$^{-1}$) |
|---|---|---|---|---|---|
| 11/10/2006 | South | WAC | 4 | 2 | 5 |
| 14/07/2006 | South | NAC | 0.7 | < 7 | < 7 |
| 03/12/2008 | South | WAC | 8 | 4 | 9 |
| 14/06/2013 | North | NAC | 1 | < 10 | < 10 |
| 14/06/2013 | North | WAC | 9 | 5 | 10 |
| 25/06/2013 | North | WAC | 10 | 5 | 11 |

# Appendix

**Table A1:** Zonal velocities and their standard deviation for the North Polar Region as a function of latitudes.

| Planetocentric Latitude (Degrees) | Planetographic latitude (Degrees) | Zonal velocity u (ms$^{-1}$) | Standard deviation $\sigma_u$ (ms$^{-1}$) |
|---|---|---|---|
| 89.9 | 89.91 | 4.48 | 0.94 |
| 89.8 | 89.83 | 13.86 | 12.33 |
| 89.7 | 89.75 | 34.39 | 10.05 |
| 89.6 | 89.67 | 67.09 | 45.13 |
| 89.5 | 89.59 | 88.87 | 43.15 |
| 89.4 | 89.51 | 104.38 | 35.39 |
| 89.3 | 89.43 | 122.98 | 15.07 |
| 89.2 | 89.34 | 128.58 | 17.40 |
| 89.1 | 89.26 | 126.22 | 23.71 |
| 90.0 | 89.18 | 131.30 | 23.62 |
| 88.9 | 89.10 | 134.99 | 20.76 |
| 88.7 | 88.94 | 140.52 | 18.07 |
| 88.5 | 88.77 | 141.5 | 15.15 |
| 88.3 | 88.61 | 133.65 | 15.19 |
| 88.1 | 88.45 | 123.87 | 15.49 |
| 87.9 | 88.29 | 117.62 | 12.68 |
| 87.7 | 88.12 | 112.07 | 11.67 |
| 87.5 | 87.96 | 106.79 | 8.61 |
| 87.3 | 87.80 | 105.39 | 8.03 |
| 87.1 | 87.63 | 102.84 | 8.2 |
| 86.9 | 87.47 | 101.02 | 7.59 |
| 86.7 | 87.31 | 100.11 | 7.32 |
| 86.5 | 87.15 | 98.61 | 7.03 |
| 86.3 | 86.98 | 96.32 | 6.85 |
| 86.1 | 86.82 | 94.85 | 6.8 |
| 85.9 | 86.66 | 93.38 | 7.21 |
| 85.7 | 86.49 | 91.45 | 7.22 |
| 85.5 | 86.33 | 88.36 | 7.46 |
| 85.3 | 86.17 | 85.16 | 7.31 |
| 85.1 | 86.00 | 80.42 | 8.33 |
| 84.9 | 85.84 | 75.21 | 8.56 |
| 84.7 | 85.68 | 70.55 | 8.92 |
| 84.5 | 85.52 | 65.69 | 8.76 |
| 84.3 | 85.35 | 60.37 | 9.39 |
| 84.1 | 85.19 | 54.78 | 8.47 |
| 83.9 | 85.03 | 51.05 | 7.92 |
| 83.7 | 84.86 | 47.99 | 7.44 |
| 83.5 | 84.70 | 45.37 | 6.98 |
| 83.3 | 84.54 | 43.45 | 7.07 |
| 83.1 | 84.37 | 41.22 | 7.38 |
| 82.9 | 84.21 | 38.7 | 7.34 |
| 82.7 | 84.04 | 37.25 | 7.82 |
| 82.5 | 83.88 | 35.54 | 8.28 |
| 82.3 | 83.72 | 32.6 | 8.53 |
| 82.1 | 83.55 | 30.43 | 8.79 |
| 81.9 | 83.39 | 28 | 8.87 |
| 81.7 | 83.23 | 25.61 | 8.77 |
| 81.5 | 83.06 | 23.14 | 8.56 |
| 81.3 | 82.90 | 20.07 | 7.75 |
| 81.1 | 82.73 | 16.7 | 7.56 |
| 80.9 | 82.57 | 14.14 | 7.38 |
| 80.7 | 82.41 | 12.65 | 7.08 |
| 80.5 | 82.24 | 10.18 | 7.23 |
| 80.3 | 82.08 | 9.42 | 7.37 |
| 80.1 | 81.91 | 9.26 | 7.34 |
| 79.9 | 81.75 | 7.91 | 7.86 |
| 79.7 | 81.58 | 7.88 | 8.25 |
| 79.5 | 81.42 | 8.24 | 8.65 |
| 79.3 | 81.25 | 8.72 | 8.8 |
| 79.1 | 81.09 | 10.21 | 8.34 |
| 78.9 | 80.93 | 11.28 | 8.26 |
| 78.7 | 80.76 | 12.46 | 8.56 |
| 78.5 | 80.60 | 14.97 | 9.15 |
| 78.3 | 80.43 | 16.94 | 9.89 |
| 78.1 | 80.27 | 18.46 | 10.21 |
| 77.9 | 80.10 | 22.93 | 11.44 |
| 77.7 | 79.93 | 27.04 | 12.05 |
| 77.5 | 79.77 | 29.73 | 13.29 |
| 77.3 | 79.60 | 37.52 | 18.48 |
| 77.1 | 79.44 | 49.05 | 24.61 |
| 76.9 | 79.27 | 59.13 | 28.35 |
| 76.7 | 79.11 | 70.3 | 29.94 |
| 76.5 | 78.94 | 80.58 | 28.05 |
| 76.3 | 78.78 | 88.46 | 23.9 |
| 76.1 | 78.61 | 95.51 | 18.8 |
| 75.9 | 78.44 | 99.22 | 15.39 |
| 75.7 | 78.28 | 98.21 | 14.5 |
| 75.5 | 78.11 | 96.06 | 14.69 |
| 75.3 | 77.95 | 91.91 | 17.59 |
| 75.1 | 77.78 | 85.06 | 19.94 |
| 74.9 | 77.61 | 76.32 | 21.99 |
| 74.7 | 77.45 | 67.74 | 21.81 |
| 74.5 | 77.28 | 57.69 | 21.24 |
| 74.3 | 77.11 | 49.4 | 18.99 |
| 74.1 | 76.95 | 42.17 | 17.73 |
| 73.9 | 76.78 | 34.7 | 15.72 |
| 73.7 | 76.61 | 29.15 | 13.43 |
| 73.5 | 76.44 | 23.74 | 12.25 |
| 73.3 | 76.28 | 18.75 | 11.04 |
| 73.1 | 76.11 | 15.75 | 10.09 |
| 72.9 | 75.94 | 13.17 | 9.69 |
| 72.7 | 75.77 | 10.01 | 9.6 |
| 72.5 | 75.61 | 7.99 | 8.7 |
| 72.3 | 75.44 | 6.15 | 9.03 |
| 72.1 | 75.27 | 4.09 | 8.67 |
| 71.9 | 75.10 | 2.55 | 8.45 |
| 71.7 | 74.93 | 0.53 | 8.18 |
| 71.5 | 74.77 | -0.81 | 7.72 |
| 71.3 | 74.60 | -1.79 | 7.62 |
| 71.1 | 74.43 | -2.81 | 7.36 |
| 70.9 | 74.26 | -3.53 | 7.32 |
| 70.7 | 74.09 | -3.35 | 7.37 |
| 70.5 | 73.92 | -3.48 | 7.07 |
| 70.3 | 73.75 | -4.2 | 6.96 |
| 70.1 | 73.58 | -4.36 | 6.79 |
| 69.9 | 73.41 | -4.3 | 7.04 |
| 69.7 | 73.24 | -4.89 | 7.43 |
| 69.5 | 73.07 | -5.69 | 7.42 |
| 69.3 | 72.90 | -5.31 | 7.47 |
| 69.1 | 72.73 | -4.74 | 7.16 |
| 68.9 | 72.56 | -4.68 | 7.25 |
| 68.7 | 72.39 | -4.19 | 7.43 |
| 68.5 | 72.22 | -3.79 | 6.85 |
| 68.3 | 72.05 | -4.4 | 7.72 |
| 68.1 | 71.88 | -4.42 | 7.99 |
| 67.9 | 71.71 | -3.74 | 6.77 |
| 67.7 | 71.54 | -3.19 | 6.94 |
| 67.5 | 71.37 | -2.32 | 7.5 |
| 67.3 | 71.20 | -1.63 | 6.86 |
| 67.1 | 71.03 | 0.2 | 6.88 |
| 66.9 | 70.86 | 0.83 | 7.36 |
| 66.7 | 70.68 | 0.86 | 6.29 |
| 66.5 | 70.51 | 0.08 | 6.63 |
| 66.3 | 70.34 | -0.3 | 7.78 |
| 66.1 | 70.17 | 1.84 | 7.49 |
| 65.9 | 69.99 | 4.23 | 5.61 |
| 65.7 | 69.82 | 5.05 | 6.09 |
| 65.5 | 69.65 | 7.01 | 6.26 |
| 65.3 | 69.48 | 9.16 | 7.28 |
| 65.1 | 69.30 | 16.2 | 6.68 |
| 64.9 | 69.13 | 17.74 | 5.73 |
| 64.7 | 68.96 | 21.63 | 5.35 |
| 64.5 | 68.78 | 23.74 | 6.83 |
| 64.3 | 68.61 | 26.18 | 6.56 |
| 64.1 | 68.44 | 27.39 | 6.61 |
| 63.9 | 68.26 | 33.02 | 3.93 |
| 63.7 | 68.09 | 32.23 | 3.69 |
| 63.5 | 67.91 | 36 | 5.32 |
| 63.3 | 67.74 | 37.57 | 5.98 |
| 63.1 | 67.56 | 45.19 | 0.21 |

**Table A2:** Zonal velocities and their standard deviation in the South Polar Region as a function of latitudes after longitudinal binning.

| Planetocentric Latitude (Degrees) | Planetographic latitude (Degrees) | Zonal velocity u (ms$^{-1}$) | Standard deviation $\sigma_u$ (ms$^{-1}$) |
|---|---|---|---|
| -89.5 | -89.59 | 69.55 | 0.00 |
| -89.4 | -89.51 | 75.57 | 8.72 |
| -89.3 | -89.43 | 86.42 | 10.24 |
| -89.2 | -89.34 | 91.65 | 6.98 |
| -89.1 | -89.26 | 100.40 | 9.93 |
| -89.0 | -89.18 | 100.40 | 9.93 |
| -88.9 | -89.10 | 115.0 | 18.68 |
| -88.6 | -88.86 | 151.84 | 14.42 |
| -88.3 | -88.61 | 157.46 | 12.25 |
| -88 | -88.37 | 152.51 | 7.45 |
| -87.7 | -88.12 | 143.07 | 11.03 |
| -87.4 | -87.88 | 142.52 | 12.15 |
| -87.1 | -87.63 | 141.18 | 11.81 |
| -86.8 | -87.39 | 135.92 | 13.63 |
| -86.5 | -87.15 | 135.42 | 14.65 |
| -86.2 | -86.90 | 138.11 | 15.87 |
| -85.9 | -86.66 | 134.27 | 17.13 |
| -85.6 | -86.41 | 128.14 | 14.01 |
| -85.3 | -86.17 | 126.14 | 14.25 |
| -85 | -85.92 | 117.45 | 19.93 |
| -84.7 | -85.68 | 108.22 | 18.58 |
| -84.4 | -85.43 | 97.2 | 20.42 |
| -84.1 | -85.19 | 85.12 | 14.32 |
| -83.8 | -84.94 | 82.84 | 13.62 |
| -83.5 | -84.70 | 77.03 | 13.52 |
| -83.2 | -84.45 | 72.1 | 12.78 |
| -82.9 | -84.21 | 65.46 | 14.35 |
| -82.6 | -83.96 | 56.93 | 12.35 |
| -82.3 | -83.72 | 51.27 | 11.86 |
| -82 | -83.47 | 45.55 | 11.04 |
| -81.7 | -83.23 | 41.96 | 9.87 |
| -81.4 | -82.98 | 38.31 | 10.16 |
| -81.1 | -82.73 | 35.22 | 9.76 |
| -80.8 | -82.49 | 31.55 | 9.57 |
| -80.5 | -82.24 | 27.19 | 9.46 |
| -80.2 | -81.99 | 22.68 | 9.86 |
| -79.9 | -81.75 | 18.98 | 9.14 |
| -79.6 | -81.50 | 15.98 | 7.75 |
| -79.3 | -81.25 | 12.5 | 7.71 |
| -79 | -81.01 | 10.75 | 7.36 |
| -78.7 | -80.76 | 10.09 | 7.21 |
| -78.4 | -80.51 | 8.53 | 8.38 |
| -78.1 | -80.27 | 6.43 | 8.55 |
| -77.8 | -80.02 | 6.08 | 8.4 |
| -77.5 | -79.77 | 5.64 | 8.7 |
| -77.2 | -79.52 | 4.17 | 8.66 |
| -76.9 | -79.27 | 4.73 | 8.36 |
| -76.6 | -79.02 | 6.09 | 8.83 |
| -76.3 | -78.78 | 5.01 | 8.56 |
| -76 | -78.53 | 4.67 | 7.99 |
| -75.7 | -78.28 | 5.29 | 7.72 |
| -75.4 | -78.03 | 5.32 | 7.81 |
| -75.1 | -77.78 | 7.57 | 8.45 |
| -74.8 | -77.53 | 8.08 | 7.48 |
| -74.5 | -77.28 | 7.93 | 8.04 |
| -74.2 | -77.03 | 9.72 | 8.14 |
| -73.9 | -76.78 | 11.49 | 8.47 |
| -73.6 | -76.53 | 13.68 | 9.05 |
| -73.3 | -76.28 | 15.61 | 9.51 |
| -73 | -76.03 | 19.79 | 11.05 |
| -72.7 | -75.77 | 24.56 | 11.92 |
| -72.4 | -75.52 | 29.04 | 14.19 |
| -72.1 | -75.27 | 37.87 | 16.7 |
| -71.8 | -75.02 | 47.91 | 19.5 |
| -71.5 | -74.77 | 65.39 | 23.64 |
| -71.2 | -74.51 | 78.71 | 19.15 |
| -70.9 | -74.26 | 85.23 | 15.07 |
| -70.6 | -74.01 | 88.38 | 12.49 |
| -70.3 | -73.75 | 86.49 | 12.63 |
| -70 | -73.50 | 83.64 | 13.8 |
| -69.7 | -73.24 | 75.82 | 15.82 |
| -69.4 | -72.99 | 67.31 | 17 |
| -69.1 | -72.73 | 56.84 | 19 |
| -68.8 | -72.48 | 43.29 | 17.41 |
| -68.5 | -72.22 | 30.34 | 13.18 |
| -68.2 | -71.97 | 23.96 | 12.63 |
| -67.9 | -71.71 | 19.17 | 13.03 |
| -67.6 | -71.46 | 14.41 | 12.06 |
| -67.3 | -71.20 | 10.2 | 11.57 |
| -67 | -70.94 | 6.59 | 12.91 |
| -66.7 | -70.68 | 0.51 | 13.77 |
| -66.4 | -70.43 | -4.32 | 11.6 |
| -66.1 | -70.17 | -6.07 | 10.83 |
| -65.8 | -69.91 | -9.27 | 10.83 |
| -65.5 | -69.65 | -11.41 | 9.94 |
| -65.2 | -69.39 | -12.38 | 11.38 |
| -64.9 | -69.13 | -13.89 | 11.98 |
| -64.6 | -68.87 | -14.16 | 11.54 |
| -64.3 | -68.61 | -14.61 | 11.24 |
| -64 | -68.35 | -15.32 | 10.41 |
| -63.7 | -68.09 | -15.39 | 9.32 |
| -63.4 | -67.83 | -15.55 | 9.07 |
| -63.1 | -67.56 | -14.82 | 8.79 |